\documentclass{aa}  
\usepackage[utf8]{inputenc}
\usepackage{booktabs}
\usepackage{amsmath}
\usepackage{amssymb}
\usepackage{graphicx}
\usepackage{siunitx} 
\usepackage{textcomp} 
\usepackage{hyperref}
\usepackage{verbatim}
\usepackage{comment}

\usepackage{orcidlink}
\usepackage{xspace}

\usepackage{txfonts}
\newcommand{\source}{GS~1843+00\xspace}

\newcommand{\rxte}{\textit {RXTE}}
\newcommand{\nustar}{\textit {NuSTAR}\xspace}
\newcommand{\chandra}{\textit {Chandra}}

\newcommand{\meer}{{MeerKAT}}

\DeclareUnicodeCharacter{2212}{-}
\begin{document} 

 \nolinenumbers
 \let\linenumbers\relax

\title{\textbf{Reawakening of GS~1843+00: A Multiwavelength Study of its 2026 Giant X-ray Outburst}}
   \titlerunning{GS~1843+00}
    \authorrunning{G.~K.~Jaisawal et al.}

\author{ Gaurava~K.~Jaisawal\orcidlink{0000-0002-6789-2723}\inst{1}
        \and Georgios Vasilopoulos\orcidlink{0000-0003-3902-3915}\inst{2}
       \and Francesco Carotenuto\orcidlink{0000-0002-0426-3276}\inst{3}
       \and Hannah C. I. Wichern\orcidlink{0009-0004-1442-619X}\inst{1}
        }
\institute{
  DTU Space, Technical University of Denmark, \text{\O}rsteds Plads 348, DK-2800 Lyngby, Denmark
    \email{gaurava@space.dtu.dk}
  \and Department of Physics, National and Kapodistrian University of Athens, University Campus Zografos, GR 15784, Athens, Greece
  \and INAF-Osservatorio Astronomico di Roma, Via Frascati 33, I-00078, Monte Porzio Catone (RM), Italy
}

   \date{Received; accepted }

 \abstract{}{}{}{}{} 
 

\abstract
{GS~1843+00 (Ginga~1843+009) is a poorly studied Be/X-ray binary system that underwent a major X-ray outburst in 2026 February--May after nearly sixteen years of quiescence.}
{We present an X-ray, optical, and radio study of GS~1843+00 during this outburst, focusing on the evolution of the neutron star emission geometry, Be-disk connection, and possible outflow activity.}
{We analyse three \nustar observations obtained at different stages of the outburst, complemented by archival \rxte{} and \chandra{} data, optical spectroscopy and imaging with the Nordic Optical Telescope, and a MeerKAT radio observation.}
{\nustar observations reveal coherent X-ray pulsations around 29.7~s with a spin-up rate of $(-5.1\pm0.1)\times10^{-8}$~s~s$^{-1}$ during the outburst. The pulse profiles show strong evolution with energy and outburst phase, with distinct morphologies even at comparable luminosities. The pulsed fraction varies between 1 and 16\% and exhibits a complex reversal in energy dependence above 7~keV, increasing with energy during the first two observations and decreasing during the third. The 3--79~keV broadband spectra are described by thermal Comptonization with a blackbody and an Fe~K$\alpha$ emission line, with no evidence for a cyclotron absorption feature. Phase-resolved spectroscopy reveals variations in the Comptonization and blackbody parameters, indicating changes in the accretion-column structure. Optical spectroscopy confirms the presence of a Be-circumstellar disk through H$\alpha$ and Paschen-series emission lines. No radio counterpart is detected, with a $3\sigma$ upper limit of 36~$\mu$Jy~beam$^{-1}$ at 1.28~GHz.}
{The observed pulse profile evolution, spectral variability, and reversal of the pulsed-fraction energy dependence suggest substantial changes in the accretion-column geometry during the outburst. Weaker H$\alpha$ emission compared with previous epochs indicates that the Be-circumstellar disk was in a different state during the current episode, while the radio non-detection suggests that any associated outflow activity was faint or below the detection limit. Together, these results reveal a complex interplay between Be-disk evolution, outflows, and neutron-star accretion geometry in GS~1843+00.}


   \maketitle

\section{Introduction}
Be/X-ray binaries (BeXRBs) constitute a major subclass of the high-mass X-ray binary population \citep{Reig2011}. These systems typically consist of a magnetized neutron star orbiting a 10--20~M$_{\odot}$ Be star. The massive optical companion is a non-supergiant OB-type star characterized by hydrogen emission lines in its optical and infrared spectra. These emission features arise from an equatorial circumstellar disk driven by the rapid rotation of the Be star \citep{Porter2003}. BeXRBs generally have wide, moderately eccentric orbits that allow the neutron star to interact periodically with the Be-circumstellar disk. Such interactions lead to transient X-ray activity through episodic mass transfer onto the neutron star. Variations in the density, size, and orientation of the Be disk can significantly influence the accretion process and the resulting X-ray outbursts \citep{Negueruela1998, 2019MNRAS.488..387B}.

Two classes of X-ray outbursts are observed in BeXRBs \citep{Stella1986, Negueruela1998, Reig2011}. Normal (Type~I) outbursts occur quasi-periodically around periastron passage, typically lasting 20--30\% of the orbital period with peak luminosities of $10^{36-37}$~erg~s$^{-1}$.  Giant (Type~II) outbursts, by contrast, are less frequent, are not restricted to periastron passage, and can persist for weeks to months.
They reach peak luminosities of $10^{37-38}$~erg~s$^{-1}$, corresponding to increases of four to five orders of magnitude above quiescence (e.g., \citealt{2016MNRAS.461L..97J, 2018ApJ...863....9W, 2020MNRAS.494.5350V}). These events are thought to result from episodes of enhanced mass accretion caused by large-scale changes in the Be-circumstellar disk, including disk expansion, warping, and tilting \citep{Negueruela1998, 2013PASJ...65...83M, 2013PASJ...65...41O, 2019MNRAS.488..387B, 2023MNRAS.518.5089C}.

The strong magnetic field of the neutron star channels accreted matter onto the magnetic poles, forming hot spots or accretion columns that result in coherent X-ray pulsations at its spin period \citep{Nagase1989}. The observed pulsed profiles are governed by the geometry of the surface emitting regions and radiative processes within the column \citep{Basko1975, Becker2007}. Broadband X-ray spectra of accreting pulsars are generally described by a hard continuum with a high-energy exponential cutoff below 35~keV. Many systems also exhibit cyclotron resonance scattering features (CRSFs) at 10--100~keV, produced by the resonant scattering of photons by electrons in quantized Landau states \citep{Jaisawal2017, Staubert2007A&A...465L..25S}. These features provide a direct estimate of the neutron star magnetic field strength, typically of the order of $\sim$10$^{12}$~G. Fluorescent iron emission lines are commonly observed, arising from X-ray irradiation of relatively cool material in the accretion disk, surroundings, or stellar wind (e.g., \citealt{2019ApJ...885...18J}). Studying spectral and timing properties provides powerful diagnostics of the accretion environment and emission geometry of X-ray pulsars, while their evolution across the critical luminosity offers insights into transitions between different mass accretional regimes \citep{Becker2007, 2015MNRAS.447.1847M}.

GS~1843+00 (Ginga~1843+009) is a BeXRB hosting a neutron star with a spin period of 29.5~s. The source was discovered on 1988 April 3 during a Galactic plane scan of the Scutum region by the \textit{Ginga} observatory \citep{1988IAUC.4583....1M, 1988IAUC.4587....1M}. A follow-up \textit{Ginga} observation on 1988 April 19 confirmed coherent pulsations and revealed an unusually low pulsed fraction of $\approx$4\% in the 2--37~keV band \citep{1990ApJ...356L..47K}. 

Nearly nine years later, the source reappeared during an outburst on 1997 March 3 by the Burst and Transient Source Experiment (BATSE) aboard the \textit{Compton Gamma Ray Observatory} \citep{1997IAUC.6586....2W}. BATSE detected clear pulsations at 29.563~s with a pulsed fraction of $\approx$7\% in the 20--50~keV range \citep{1997IAUC.6586....2W}. A subsequent \textit{RXTE} observation on 1997 March 5 confirmed pulsations in the 2--60~keV band \citep{1997IAUC.6595....2T}. Moreover, a \textit{BeppoSAX} observation on 1997 April 4 measured a spin period of 29.477~s, corresponding to a spin-up rate of $\dot{P} = (-3.79 \pm 0.10) \times 10^{-8}$~s~s$^{-1}$ during the outburst \citep{2000A&A...357..501P}. During this observation, the pulse profile was singly peaked at higher energies, while the pulsed fraction decreased from 15\% to 5\% across the 2--60~keV range. The broadband 0.3--100~keV spectrum was described by an absorbed cutoff power-law model at a flux level of (2.9$\pm$0.3)$\times$10$^{-9}$~erg~cm$^{-2}$~s$^{-1}$, without any evidence for a CRSF  \citep{2000A&A...357..501P}.

Using the improved X-ray position from \textit{ROSAT} \citep{1997IAUC.6645....2D}, optical and infrared follow-up observations identified a faint ($V = 20.9$~mag), heavily reddened early-type counterpart of \source\ exhibiting Balmer and Paschen emission lines and an infrared excess \citep{2001A&A...371.1018I}. The companion was classified as a B0--B2~IV--Ve star, confirming the system as a BeXRB. The distance was suggested to be $\gtrsim$10~kpc \citep{2001A&A...371.1018I}, while the Gaia DR3 parallax \citep{2021AJ....161..147B} corresponds to a distance of $5.1^{+4.2}_{-1.9}$~kpc. 
In our study, we use 5.1 and 10~kpc as reference distances for comparison. 

Several weaker X-ray outbursts from \source were detected by \textit{INTEGRAL} and \textit{Swift}/BAT at peak fluxes of 7–50~mCrab between 2003 and 2006 \citep{2003ATel..159....1C,2005ATel..472....1L,2006ATel..769....1K,2007ESASP.622..499S}. Based on the recurrence of these events, a probable orbital period of 160~days has been suggested \citep{2007ESASP.622..499S}. The source again entered a renewed active phase in 2009 March, with a relatively stronger outburst \citep{2009ATel.1959....1K,2009ATel.1977....1R} that was followed by two weaker events in 2009 November and 2010 March \citep{2026ATel17706....1N}. Using \textit{Swift}/XRT observations on 2009 March 15, near the peak of the 2009 outburst, the X-ray position was refined to R.A. (J2000) $= 281.40372^\circ$ (18$^{\rm h}$ 45$^{\rm m}$ 36.9$^{\rm s}$) and Dec. (J2000) $= 0.86338^\circ$ (00$^\circ$ 51$^{\prime}$ 48.2$^{\prime\prime}$), with a 90\% confidence uncertainty of 1.5~arcsec \citep{2009ATel.1973....1K}. The corresponding source flux was estimated to be $5.2 \times 10^{-10}$~erg\,cm$^{-2}$\,s$^{-1}$ in the 0.3--10~keV and $1.3 \times 10^{-9}$~erg\,cm$^{-2}$\,s$^{-1}$ in the 15--150~keV band  \citep{2009ATel.1973....1K}.

After nearly 16 years of quiescence, \textit{MAXI}/GSC reported the onset of a new X-ray outburst from the direction of \source\ on 2026 February 19 (MJD~61090) \citep{2026ATel17706....1N}. A follow-up Target-of-Opportunity observation with \textit{NuSTAR} on 2026 March 10--11 confirmed renewed activity \citep{2026ATel17722....1J}. In this work, we present a comprehensive multiwavelength study of \source\ during its three-month-long outburst (2026 February--May), combining X-ray, optical, and radio observations. We perform X-ray timing and spectral analyses using three \textit{NuSTAR} observations, investigating the evolution of the pulse profile morphology and pulsed fraction across the 3--79~keV band. Phase-averaged spectroscopy complemented by phase-resolved spectroscopy are also performed to probe X-ray emission geometry. To explore the long-term evolution of the source, we study archival \textit{RXTE} observations obtained near the peaks of the 1997 March and 2009 March outbursts. In addition, a quiescent \textit{Chandra} observation from 2013 is used to refine the source position. We further present optical and near-infrared imaging and spectroscopy obtained with the Nordic Optical Telescope (NOT), together with a MeerKAT radio observation during the declining phase of the outburst.
The structure of the paper is as follows. Section~2 describes the observations and data reduction for the X-ray (\textit{NuSTAR}, \textit{RXTE}, and \chandra), optical, and radio datasets. Section~3 presents the results from X-ray timing, pulse-profile analysis, and spectroscopy, followed by the optical and radio results. Section~4 discusses the implications of our findings, and Section~5 summarizes our conclusions.

\begin{figure}
\centering
\includegraphics[height=2.7in, width=3.45in, angle=0]{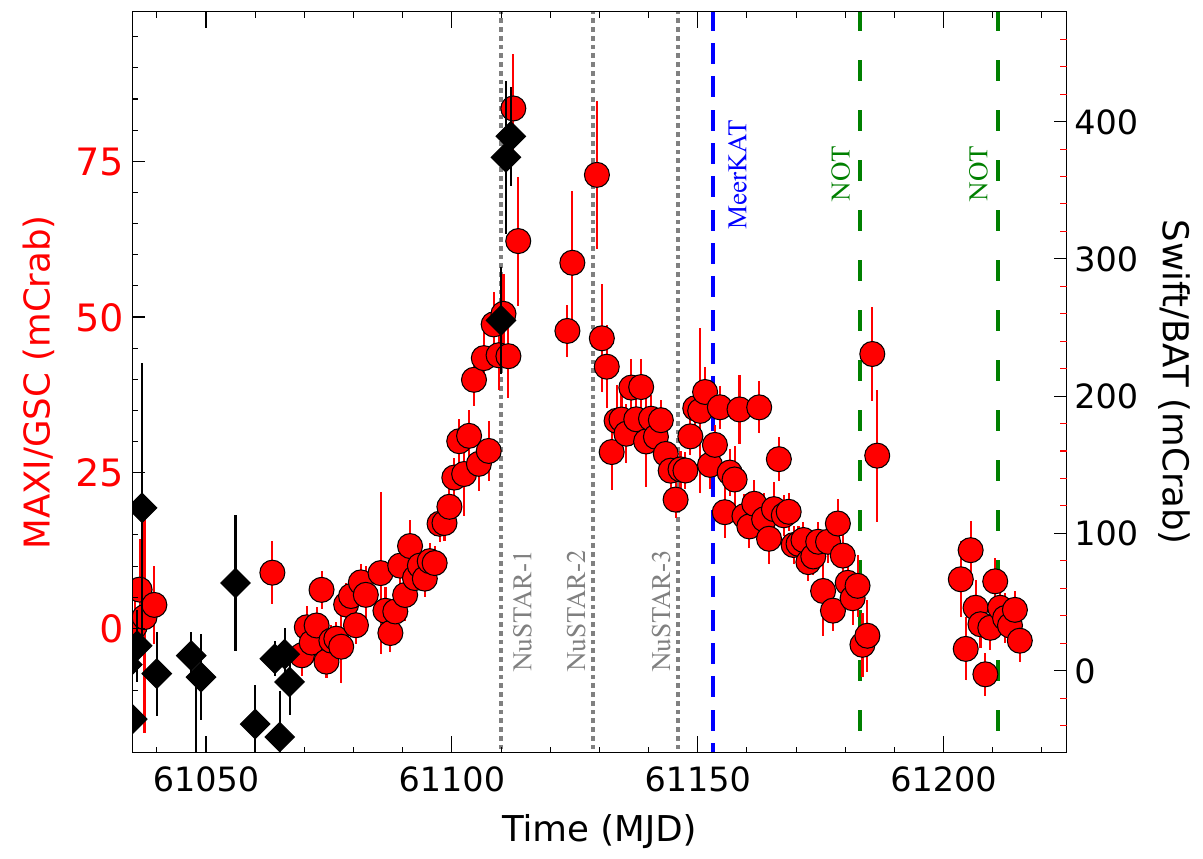}
\caption{One-day averaged monitoring light curves of \source{} using \textit{MAXI}/GSC (2–20~keV; red circles) and \textit{Swift}/BAT (15–50~keV; black diamonds). The vertical grey dotted lines indicate the start times of \textit{NuSTAR} observations. The blue and green vertical lines show the time of radio (\meer) and optical (NOT/ALFOSC) observations, respectively.}  
\label{fig:maxi-lc}
\end{figure}

\section{Observations and Data Processing}

\subsection{NuSTAR}

Following the onset of the 2026 outburst \citep{2026ATel17706....1N}, we observed \source\ with \textit{NuSTAR} \citep{2013ApJ...770..103H} on three occasions between March and April 2026 (PI Jaisawal). The first observation (ObsID 91201313002; NuSTAR-1) was performed on 2026 March 10 22:32 to March 11 20:06~UTC, with a net exposure of 36~ks during the rising phase of the outburst. At this time, the source was detected at flux levels of 50~mCrab in \textit{MAXI}/GSC (2--20~keV) and 255~mCrab in \textit{Swift}/BAT (15--50~keV), as indicated by the first grey vertical line in Fig.~\ref{fig:maxi-lc}.
The second observation (ObsID 91201319002; NuSTAR-2) was carried out on 2026 March 29 18:01 to March 30 16:45~UTC, with a net exposure of 37.9~ks, approximately 19 days after the first observation (second grey vertical line in Fig.~\ref{fig:maxi-lc}). A third observation (ObsID 91201319004; NuSTAR-3) was obtained on 2026 April 16 00:30--11:15~UTC during the declining phase of the outburst. A summary of all observations is provided in Table~\ref{table:log}.

Data reduction was performed using HEASoft~v6.35.2 and the latest calibration database (20260309). Cleaned event files were generated from unfiltered data using the standard \texttt{nupipeline} task. Light curves and spectra were extracted using \texttt{nuproducts}, adopting a circular source region of radius 120~arcsec and a nearby source-free background region of identical size. Photon arrival times were corrected to the Solar System barycenter using \texttt{barycorr} with the JPL~DE405 ephemeris. Spectra were grouped to a minimum of 30~counts per bin.

\subsection{RXTE}

Two \textit{RXTE} observations of \source\ were carried out during the 1997 and 2009 outbursts. These data are used to investigate the evolution of the pulse profile morphology and spectral properties across different outburst episodes. The first observation (ObsID 20143-06-01-00; RXTE-1) was performed on 1997 March 5 with a total exposure of 9.6~ks, while the second (ObsID 94032-11-01-00; RXTE-2) was obtained on 2009 March 13 with an exposure of 3.5~ks (Table~\ref{table:log}). Both observations were conducted near the peaks of their respective outbursts, as shown by \textit{RXTE}/ASM and \textit{Swift}/BAT light curves \citep{1997IAUC.6595....2T, 2009ATel.1973....1K}.

Data analysis was performed following standard procedures described in the \rxte{} Cook Book\footnote{\url{https://heasarc.gsfc.nasa.gov/docs/xte/recipes/cook_book.html}}. We used data from the Proportional Counter Array (PCA; \citealt{2006ApJS..163..401J}), which comprised five Proportional Counter Units (PCUs) sensitive in the 2--60~keV energy range, providing a total effective area of approximately 6500~cm$^{2}$. Good time intervals were generated using standard selection criteria applied to the available PCUs.
Source light curves were extracted in the 2--60~keV band from Standard-1 mode data with a time resolution of 0.125~s using the {\tt SAEXTRCT} task. For spectral analysis, source spectra were extracted from Standard-2 data, and the corresponding response matrices were generated using the {\tt PCARSP} task. A systematic uncertainty of 0.5\% was added to the spectra. Background light curves and spectra were produced from Standard-2 data using the PCA background model provided by the instrument team.

\subsection{Chandra}

GS~1843+00 was observed with \textit{Chandra}/Advanced CCD Imaging Spectrometer (ACIS) on 2013 June 9 (ObsID~14644) for a net exposure of 5~ks (Table~\ref{table:log}). It was performed during an X-ray quiescence state of the pulsar, at a flux level of 10$^{-13}$~erg~cm$^{-2}$~s$^{-1}$ \citep{2017MNRAS.470..126T}. We processed \textit{Chandra} data with {\tt CIAO}~v4.18 \citep{2006SPIE.6270E..1VF} in the presence of calibration files using {\tt chandra\_repro} pipeline. Standard filtering and analysis procedures were applied to the ACIS data. 
A faint X-ray source was detected at the position of GS~1843+00 with a detection significance of 3.7$\sigma$. The source position was determined to be R.A. (J2000) $=281.40350^\circ$ (18$^{\rm h}$45$^{\rm m}$36.84$^{\rm s}$) and Dec. (J2000) $=0.86324^\circ$ (+00$^\circ$51$'$47.7$''$), with uncertainties of about 0.12$''$ and 0.09$''$ in right ascension and declination, respectively. 
The \textit{Chandra} position significantly improves upon the previously reported \textit{Swift}/XRT localization obtained during the 2009 outburst \citep{2009ATel.1973....1K}. It is also consistent with the Gaia~DR3 counterpart (Gaia~DR3~4278536022438800640) at position of R.A. (J2000) $=281.40347^\circ$ and Dec. (J2000) $=0.86316^\circ$ within 0.3$''$, further confirming the identification of the optical counterpart of GS~1843+00 by \citet{2001A&A...371.1018I}.

\subsection{Nordic Optical Telescope}\label{sec:not}

The Nordic Optical Telescope (NOT) is a 2.56~m telescope located at the Roque de los Muchachos Observatory on La Palma, Spain. We obtained optical and near-infrared spectroscopic and imaging observations using the Andalucia Faint Object Spectrograph and Camera (ALFOSC) attached to the telescope. The first spectroscopic observation started on 2026 May 23 at 01:38~UTC at an airmass of 1.37. Two independent spectroscopic exposure frames were acquired, resulting in a total integration time of 5100~s. The observations were performed using ALFOSC~Grism~\#4 together with a 1.0 arcsec slit, providing wavelength coverage over approximately 3200--9600\,\AA\ at a spectral resolution of R$\sim$360.  A second spectroscopic observation was performed on 2026 June 19, starting at 23:55 UTC at an airmass of 1.34 using ALFOSC~Grism~\#4. Two exposures of 2700~s each were obtained, resulting in a total integration time of 5400~s. In addition, optical imaging observations of the field of GS~1843+00 were acquired with ALFOSC on 2026 June 20 through $V$, $R$, and $I$ filters with exposure times of 360, 200, and 100~s, respectively. A summary of optical observations is provided in Table~\ref{table:log}.

The spectroscopic and imaging data reduction was carried out using the \texttt{PyNOT}\footnote{\url{https://github.com/jkrogager/PyNOT}} software package. The spectroscopic reduction procedure included bias subtraction, flat-field correction, cosmic-ray cleaning, wavelength calibration, sky subtraction, and one-dimensional spectral extraction. Wavelength calibration was performed using arc-lamp exposures obtained at the time of science observations. The extracted one-dimensional spectra from the individual frames were subsequently combined to improve the signal-to-noise ratio. 
We performed spectroscopic flux calibration using observations of the standard star BD+33~2642 (SP1550+330) obtained during the same observing run with the identical instrumental setup. A total exposure time of 60\,s was used for the standard-star observation. The calibrated spectra were corrected for the instrumental response and converted into physical flux units of erg\,s$^{-1}$\,cm$^{-2}$\,\AA$^{-1}$. For photometry analysis, processed and astrometry-corrected images in each band were calibrated using Gaia~DR3 field stars located within $0.1^\circ$ of the source position. The photometric zero point was obtained from the median offset between the transformed Gaia magnitudes and the instrumental magnitudes of the matched stars in our analysis (see, e.g. \citealt{2026arXiv260608752J}).

\subsection{\meer}
We performed a radio observation of \source{} using the MeerKAT interferometer \citep{Camilo2018, Jonas2016} under Director's Discretionary Time (project code DDT-20260420-GJ-01; PI Jaisawal). The observation was taken on 2026 April 23, starting at 01:50 UT, with a total integration time of 1~hour at L-band (Table~\ref{table:log}). The observing setup provided a central frequency of 1.28 GHz and a total bandwidth of 875~MHz. The source PKS~J1939--6342 was used for bandpass and flux calibration, while PKS~J1911$-$2006 was adopted as the complex gain calibrator. The data reduction was performed using the \texttt{OxKAT} pipeline \citep{oxkat}, which carries out the standard flagging, calibration, and imaging procedures through the \texttt{tricolour} \citep{Hugo_2022}, \texttt{CASA} \citep{CASA_team_2022}, and \texttt{WSCLEAN} \citep{Offringa_wsclean} software packages, respectively.

\section{Results}

Fig.~\ref{fig:maxi-lc} presents the one-day averaged monitoring light curves of \source{} obtained with \textit{MAXI}/GSC (red circles; \citealt{Matsuoka2009PASJ...61..999M}) and \textit{Swift}/BAT (black diamonds; \citealt{Krimm2013}). The \textit{MAXI}/GSC light curve shows the onset of the X-ray outburst around 2026 February 19 (MJD~61090) \citep{2026ATel17706....1N}, followed by a rapid rise and gradual decay toward the pre-outburst level by the end of 2026 May (MJD~61180). The 2026 event, therefore, represents one of the longest observed outbursts from \source{}, lasting for approximately three months. The first and second \textit{NuSTAR} observations were obtained close to the outburst peak (before and after), at comparable X-ray flux levels, whereas the third observation was performed during the decay phase at a lower intensity (Fig.~\ref{fig:maxi-lc}). The 2026 outburst appears to have reached a higher intensity compared with previous outbursts observed in 1997 and 2009 (see, e.g., \citealt{2007ESASP.622..499S}).

\begin{figure}[]
\centering
\includegraphics[height=5.2in, width=3.25in, angle=0]{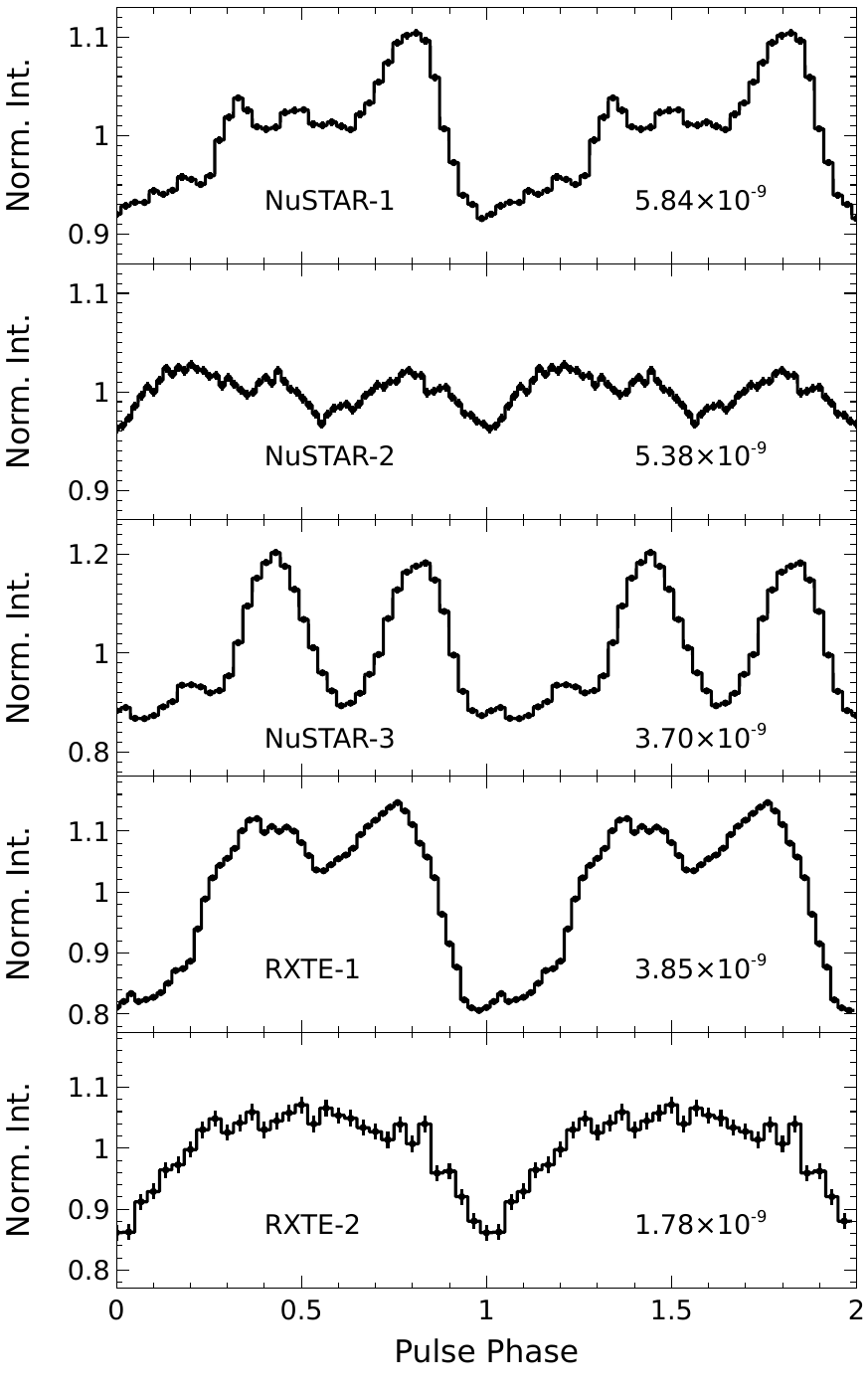}
\caption{Pulse profiles of \source obtained from \textit{NuSTAR} (3--79~keV) and \textit{RXTE} (2--60~keV) observations, shown with unabsorbed flux in the 3-80~keV range (in erg\,cm$^{-2}$\,s$^{-1}$ unit; right side of each panel). }
\label{fig:profile_broad}
\end{figure}

\begin{figure*}
\centering
\includegraphics[height=4.0in, width=2.2in, angle=0]{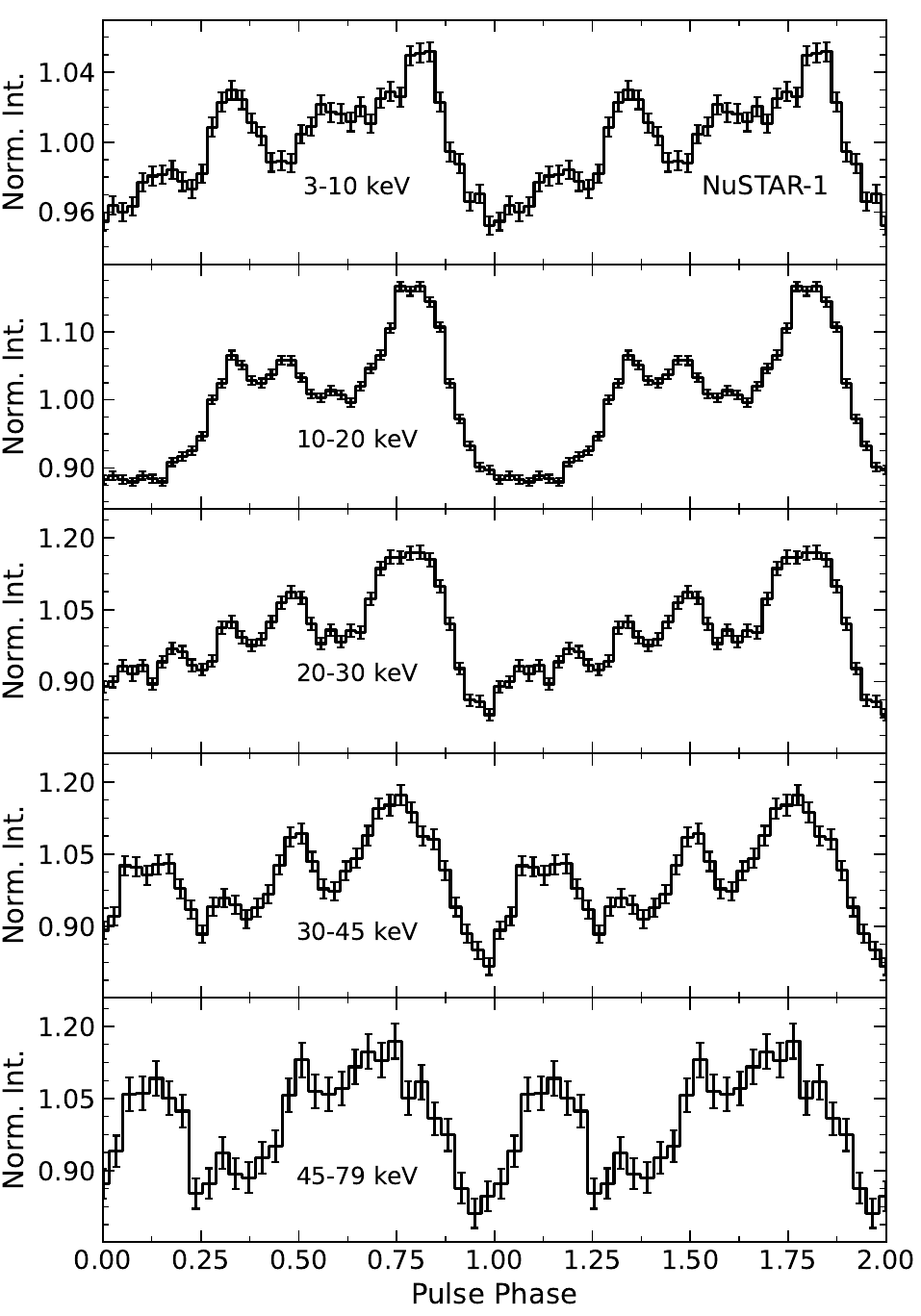}
\includegraphics[height=4.0in, width=2.2in, angle=0]{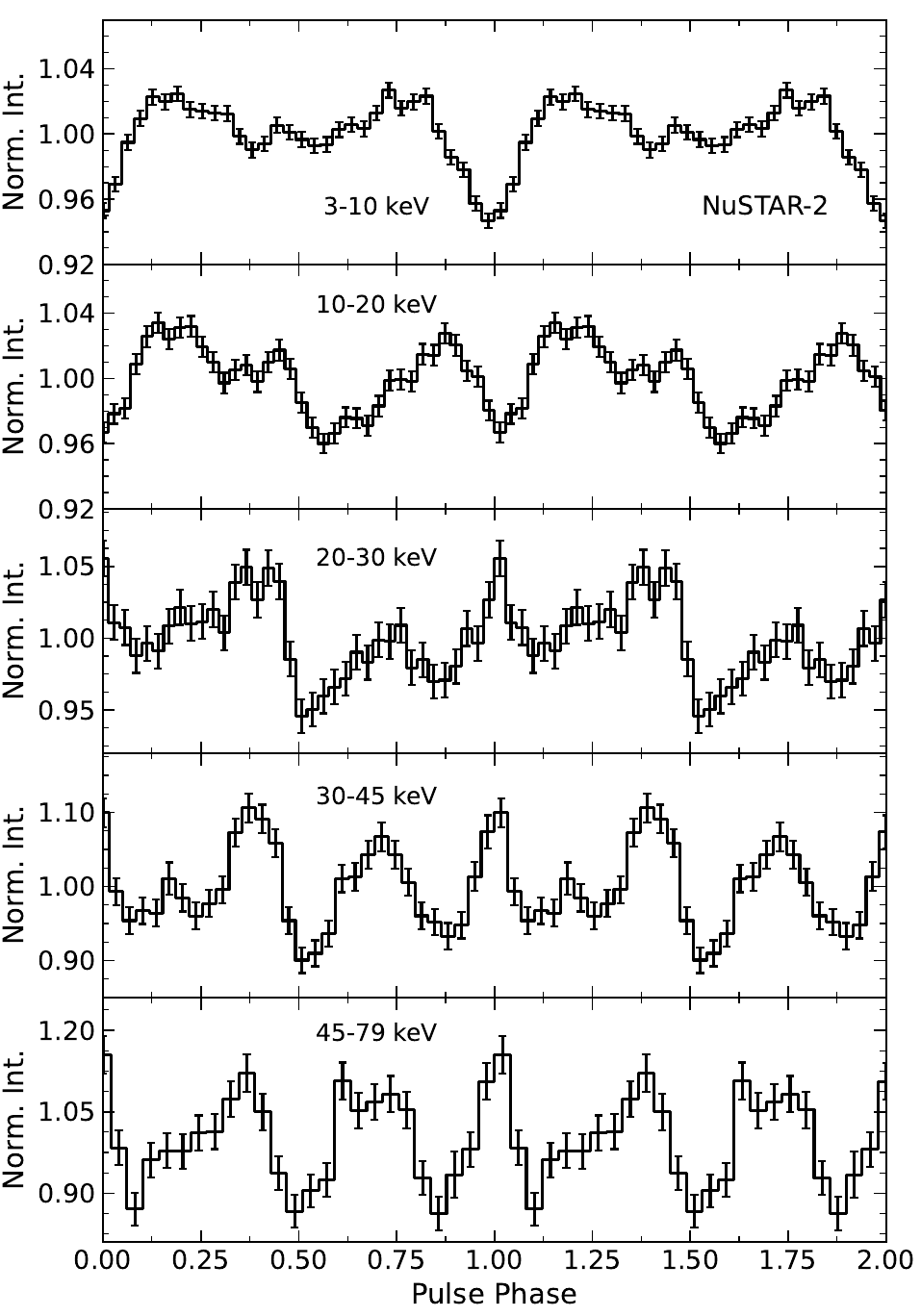} 
\includegraphics[height=4.0in, width=2.2in, angle=0]{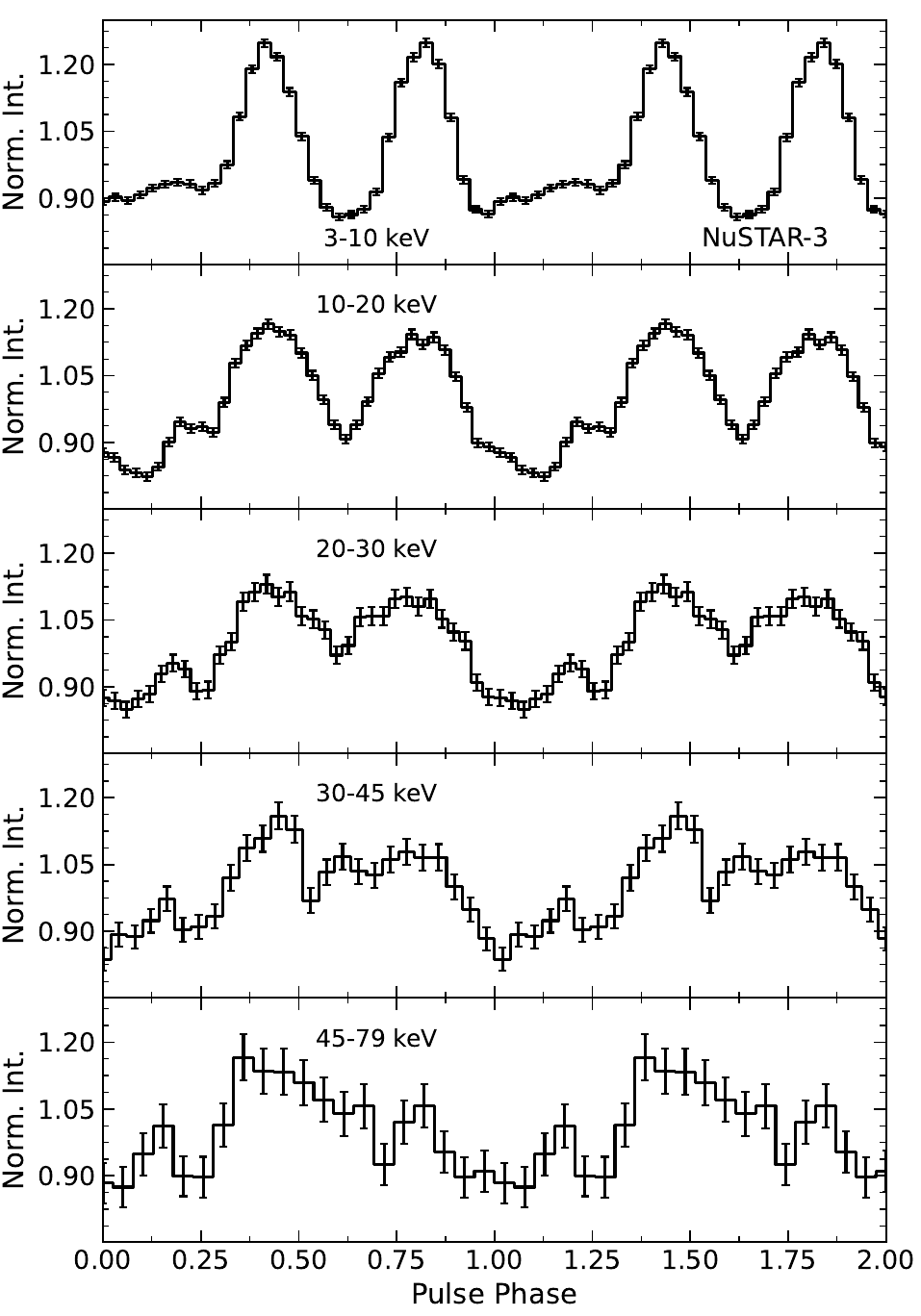} \\
\caption{Energy-resolved pulse profiles of \source{} obtained from the first (left), second (middle), and third (right) \textit{NuSTAR} observations during the 2026 outburst in 3--10, 10--20, 20--30, 30--45, and 45--79~keV energy bands. }
\label{fig:profile_energyresolved}
\end{figure*}

\begin{figure*}
\centering
\includegraphics[width=0.33\textwidth]{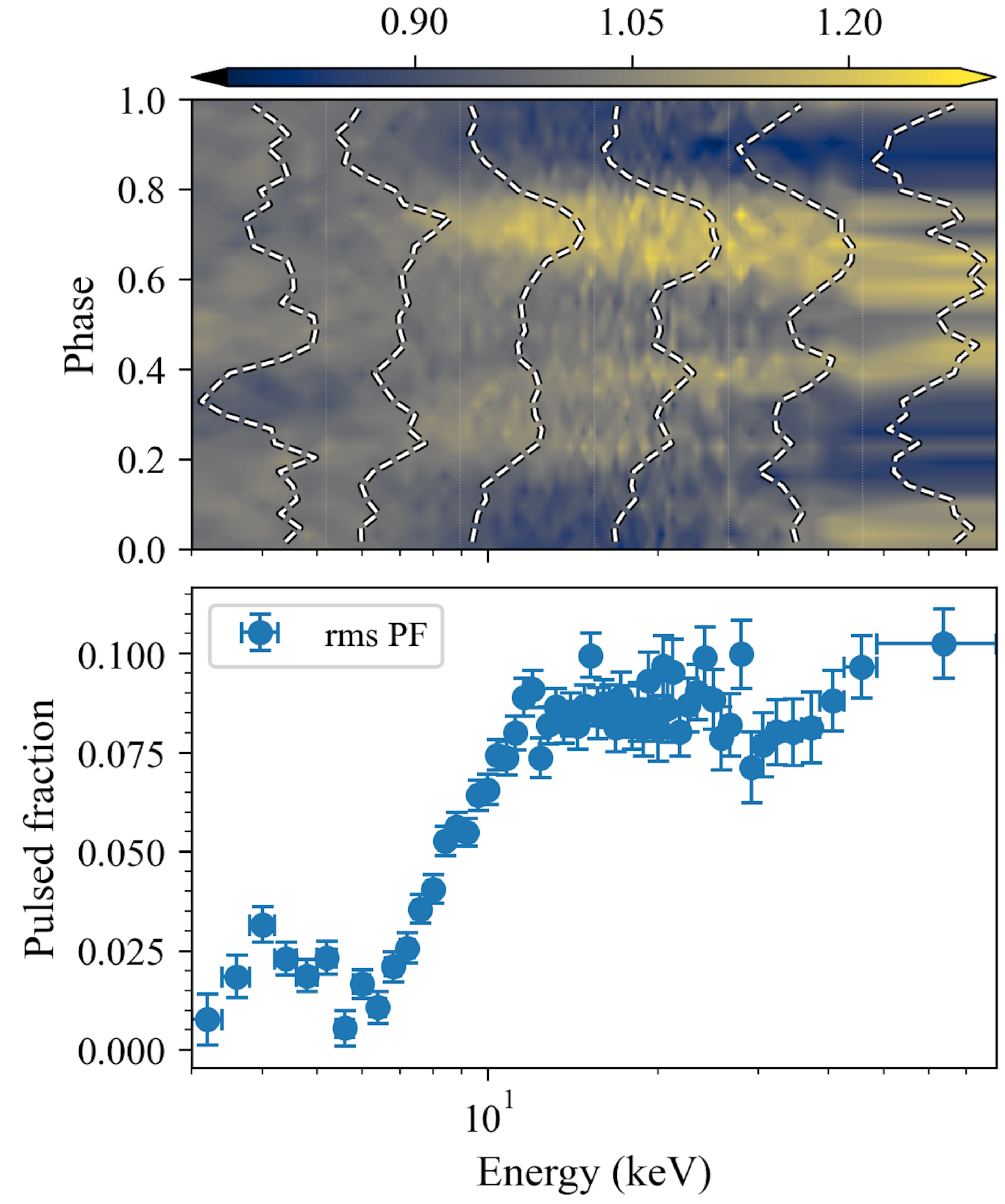}
\includegraphics[width=0.33\textwidth]{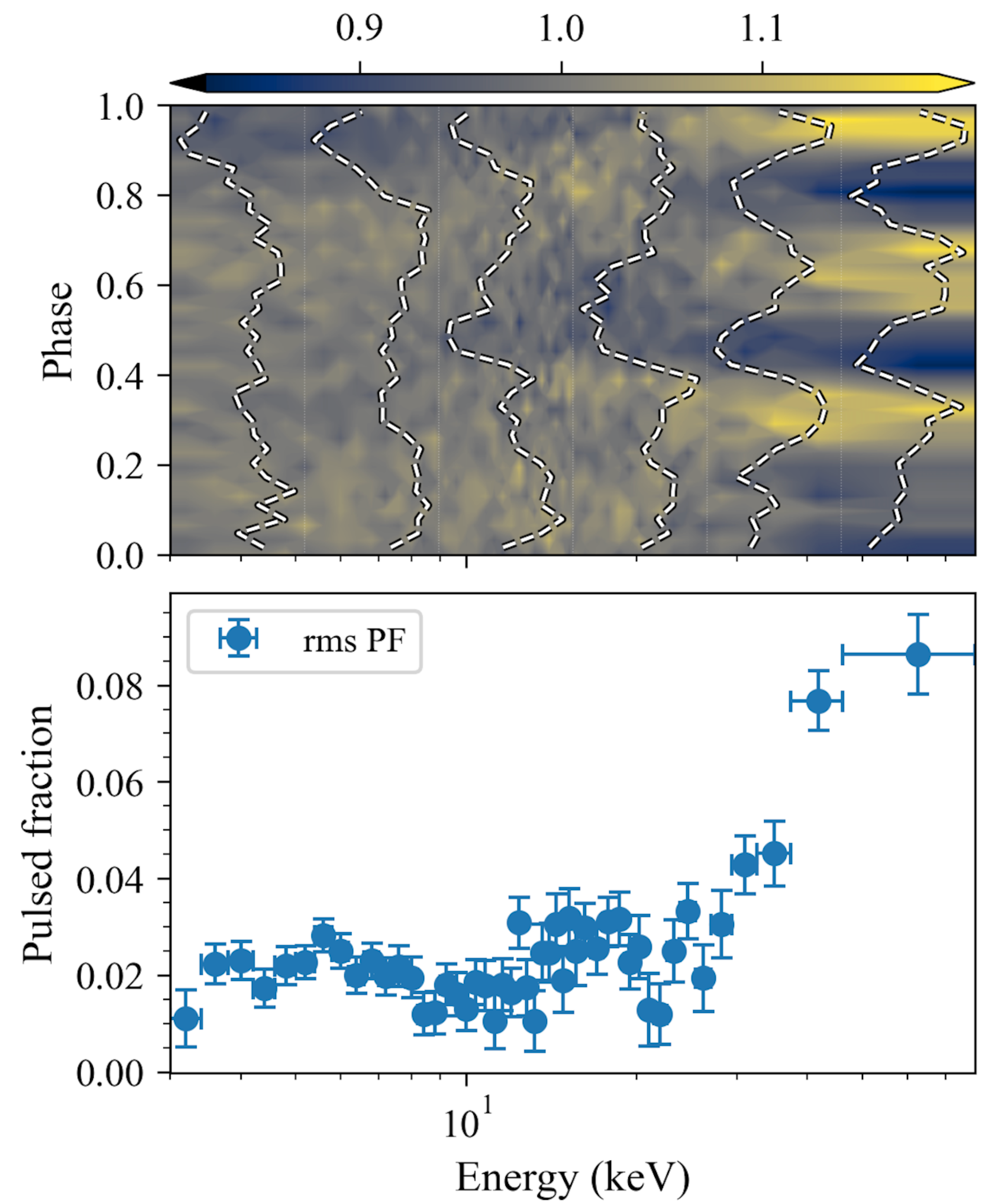} 
\includegraphics[width=0.33\textwidth]{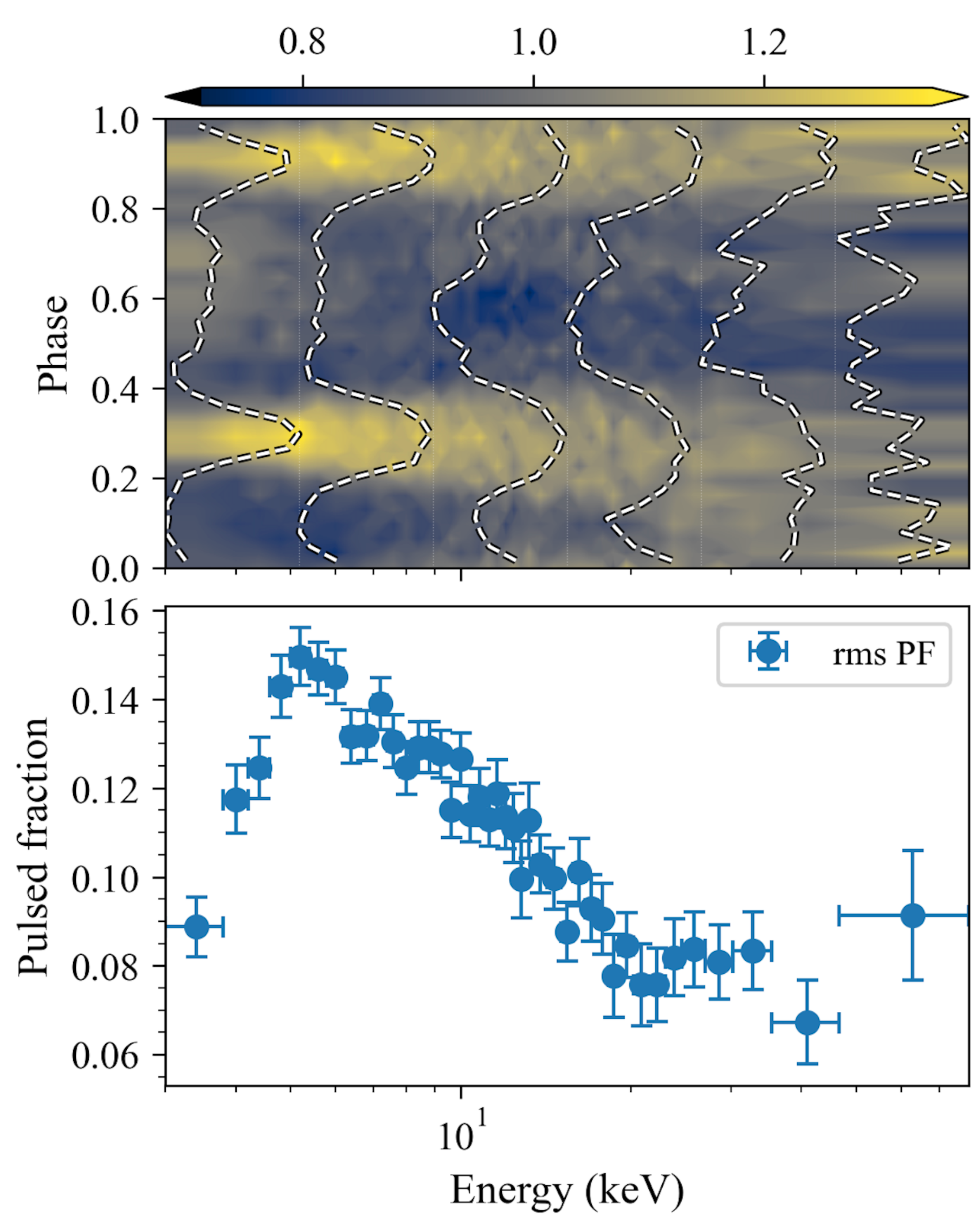} 
\caption{Energy-resolved pulse-profile heatmaps of \source{} from the first (left), second (middle), and third (right) \textit{NuSTAR} observations. Colors indicate the normalized intensity. White curves show pulse profiles in six energy bands (equally spaced in logarithmic energy), with amplitudes scaled to highlight the evolution of the pulse shape with energy. Bottom: Corresponding pulsed fractions, computed using the root-mean-square method, as a function of energy.
}
\label{fig:pf_heatmap}
\end{figure*}

\subsection{X-ray Temporal Analysis}

To study the timing properties of \source, we analysed barycenter-corrected \textit{NuSTAR} light curves from three observations obtained during the 2026 outburst.  The analysis was performed using Lomb--Scargle periodograms \citep{1976Ap&SS..39..447L} and epoch-folding techniques \citep{Leahy1987} to search for coherent pulsations from the neutron star. We analysed light curves from both focal plane modules independently, which revealed pulsations near the known spin period. To improve the signal-to-noise ratio, background-subtracted light curves from FPMA and FPMB were combined using the {\tt lcmath} tool. X-ray pulsations were clearly detected in all three observations, with periods of 29.8346$\pm$0.0008, 29.738$\pm$0.004, and 29.6802$\pm$0.0001~s for NuSTAR-1, NuSTAR-2, and NuSTAR-3, respectively  (see Fig.~\ref{fig:lspower}). We note that the strength of the pulsation signal varies between the three observations. The Leahy-normalized power spectrum shows a weaker pulsation peak during the second observation, followed by an increase in the pulsation power during the third observation (see Fig.~\ref{fig:lspower}). We estimated the period uncertainty from 5000 Monte Carlo simulations, recovering the Lomb–Scargle period for each noise-perturbed light curve. We adopt the spread of the simulated period distribution as the 1$\sigma$ error.

Using the period measurements over 34.9~day interval between the first and third \textit{NuSTAR} observations, we derive a spin-period derivative of $\dot{P} = (-5.1 \pm 0.1)\times10^{-8}$~s~s$^{-1}$. The magnitude of this period derivative is larger than that measured during the 1997 outburst \citep{2000A&A...357..501P}. 
For completeness, we also investigated the spin-period evolution of the pulsar using archival \textit{RXTE} observations. During the March 1997 outburst, the spin period was measured to be $29.5641 \pm 0.0001$~s, consistent with the value reported by \citet{1997IAUC.6595....2T}. For the March 2009 outburst, the spin period was found to be $29.526 \pm 0.001$~s from the second \textit{RXTE} observation.

Next, we folded background-subtracted light curves at the corresponding pulse periods for all observations. Fig.~\ref{fig:profile_broad} presents the broadband pulse profiles of \source\ obtained with \textit{NuSTAR} (3--79~keV) and \textit{RXTE} (2--60~keV), ordered by increasing unabsorbed flux in the 3--80~keV band based on the spectral analysis (see Section~\ref{sec:spec}). The pulse profiles show clear variations in morphology across different observations. Despite comparable flux levels between the first and second  \textit{NuSTAR} observations, the profiles differ significantly. The first observation with an unabsorbed flux of $5.84 \times 10^{-9}$~erg~cm$^{-2}$~s$^{-1}$ shows a relatively flat-topped profile followed by a narrow main peak between phases 0.65--0.90. In contrast, the second observation displays a multi-peaked structure accompanied by a lower pulsed amplitude. The third \textit{NuSTAR} observation, at an intermediate flux level of $3.7 \times 10^{-9}$~erg~cm$^{-2}$~s$^{-1}$, shows a clear double-peaked profile. Notably, the 1997 \textit{RXTE} observation has a similar intermediate flux level, but shows a less pronounced double-peaked structure.  At the lower flux of $1.78 \times 10^{-9}$~erg~cm$^{-2}$~s$^{-1}$, a single-peaked profile is detected from \rxte{} observation in 2009 (Fig.~\ref{fig:profile_broad}).

We further investigated the evolution of the averaged pulsed fraction (PF), defined as $(I_{\rm max}-I_{\rm min})/(I_{\rm max}+I_{\rm min})$, where $I_{\rm max}$ and $I_{\rm min}$ represent the maximum and minimum intensities of the pulse profile, respectively. Despite similar flux levels during the first and second \textit{NuSTAR} observations, the PF decreases from $9.5\pm0.3\%$ to $2.8\pm0.3\%$. During the third \textit{NuSTAR} observation, the PF increases to $16.5\pm0.2\%$. For comparison, the pulsed fractions measured from the 1997 and 2009 \textit{RXTE} observations are $17.2\pm0.2\%$ and $10\pm2\%$, respectively. These observations were obtained close to the respective outburst peaks (Fig.~\ref{fig:profile_broad}).

\begin{table*}
\centering
\caption{Best-fitting spectral parameters (with 90\% errors) of GS~1843+00 obtained from \nustar{} and \rxte{} observations. The model consists of an absorbed Comptonization {\tt comptt} continuum with a blackbody and a Gaussian for the iron K$\alpha$ line.}
\begin{tabular}{lccccc}
\hline
Parameters                                 & NuSTAR-1 & NuSTAR-2           &NuSTAR-3                      & RXTE-1 & RXTE-2 \\
\hline
Column density N$_{\rm H}$ (10$^{22}$ cm$^{-2}$)      & 0.9$^{+0.06}_{-0.9}$   & 0.9$^{+0.04}_{-0.9}$  & 0.9$^{+0.2}_{-0.9}$ & 4$\pm$1  & 3$\pm$2 \\

Seed photon temperature T$_{\rm 0}$ (keV) & 3.35$\pm$0.06    & 3.2$\pm$0.1    & 2.8$\pm$0.2                           & 0.4$^{+0.5}_{-0.4}$ & 1.3$\pm$0.1 \\
Plasma temperature Comp$_{\rm kT}$ (keV)  & 10.9$\pm$0.2     & 10.4$\pm$0.2    & 10.6$\pm$0.3                        & 5.9$\pm$0.1   & 6.5$\pm$0.3 \\
Optical depth $\tau$                       & 3.8$\pm$0.1      & 4.2$\pm$0.1    & 4.1$\pm$0.2                            & 11$\pm$2   & 6.4$\pm$0.4 \\
Comp$_{\rm Norm}$ (10$^{-2}$)               & 2.34$\pm$0.04   & 2.23$\pm$0.04    & 1.59$\pm$0.05                       & 4$\pm$1    & 1.8$\pm$0.1 \\
\\
Blackbody temperature BB$_{\rm kT}$ (keV)   & 1.23$\pm$0.03  & 1.25$\pm$0.02     & 1.20$\pm$0.2            & 2.1$\pm$0.3 & -- \\
Norm$_{\rm BB}$                             & 14.5$\pm$0.8   & 17.9$\pm$0.7      & 6$\pm$3           & 1.1$\pm$0.5   & --\\
Radius$_{\rm BB}$ (km)                      & 3.7$\pm$0.1    & 4.2$\pm$0.1        & 2.5$\pm$0.6          & 1.1$\pm$0.2 & -- \\
\\
Fe line energy (keV)                  & 6.40$\pm$0.02    & 6.41$\pm$0.02        & 6.39$\pm$0.09     & 6.4$\pm$0.1 & -- \\
Fe line width (keV)                  & 0.32$\pm$0.04    & 0.40$\pm$0.05        & 0.5$\pm$0.2     & 0.4$\pm$0.2 & -- \\
Line normalization (10$^{-3}$)         & 1.3$\pm$0.1     & 1.4$\pm$0.1           & 0.8$\pm$0.3          & 1.1$\pm$0.5 & -- \\
Fe line eq. width (eV)                & 120$\pm$12       & 128$\pm$10         & 118$\pm$20        & 110$\pm$45     & -- \\
\\
Flux$^a$  ($\times$10$^{-9}$)           & 5.84$\pm$0.04     & 5.38$\pm$0.05     & 3.68$\pm$0.04   & 3.85$\pm$0.07    & 1.78$\pm$0.09 \\
Luminosity$^b$ ($\times$10$^{37}$)        & $1.8^{+4.2}_{-1.1}$    & $1.7^{+3.9}_{-1.0}$   & $1.1^{+2.7}_{-0.7}$  & $1.2^{+2.8}_{-0.7}$  & $0.6^{+1.3}_{-0.4}$ \\
Luminosity$^c$  ($\times$10$^{37}$)        & 7.0$\pm$0.1           & 6.4$\pm$0.1      & 4.4$\pm$0.1   & 4.6$\pm$0.1      & 2.1$\pm$0.1 \\

$\chi^2_\nu$ ($\nu$)                  & 1.13 (2603)    & 1.10 (2579)    & 1.09 (2082)  & 1.02 (61)   & 0.90 (50) \\
\hline
\end{tabular}
\\
\flushleft
$^a$ : The 3-80 keV unabsorbed flux in unit of erg~cm$^{-2}$~s$^{-1}$.
$^b$ and $^c$ represent the corresponding luminosity in erg~s$^{-1}$, assuming distances of 5.1$_{-1.9}^{+4.2}$ and 10~kpc, respectively. 
\label{table:spec-para}
\end{table*}

\subsection{Energy-resolved pulse profiles and pulse fraction evolution with \textit{NuSTAR}}

To investigate the energy dependence of the pulse morphology, we generated energy-resolved pulse profiles in five broad bands (3--10, 10--20, 20--30, 30--45, and 45--79~keV), chosen to trace the overall evolution of the pulse shape across the \nustar bandpass. During the first observation (left side of Fig.~\ref{fig:profile_energyresolved}), the pulse profile appears complex due to multiple structures in the 3--10~keV range. One of these features evolves into a dominant peak in the 10--20~keV range between 0.65--0.90 phase. As the energy increases, the relative strengths of the observed peaks change, and the profile develops multi-peaked structures in the 20--45~keV range. At higher energies (45--79~keV), the profile becomes double-peaked, consisting of a narrow and a broad peak.

Similarly, a complex energy evolution is observed during the second observation, shown in the middle panel of Fig.~\ref{fig:profile_energyresolved}. The profile is relatively broad in the 3--10~keV band and develops a double-peaked structure between 10 and 30~keV. At higher energies, the profile evolves into a multi-peaked structure, becoming triple-peaked in the 45--79~keV band. We also observe a phase shift between the low- and high-energy pulse profiles during this observation. In particular, the intensity minimum in the 3--10~keV profile occurs at a phase offset of approximately half a cycle compared with the minima observed in the 30--45~keV and 45--79~keV profiles. During the third observation, the pulse profile shows a double-peaked structure at low energies (3--10~keV), which gradually evolves into a broader single peak at higher energies.

To quantify this evolution on a finer energy grid, we defined energy bins adaptively rather than at a fixed width. Starting from an initial grid of 0.4~keV, contiguous fine bins were merged until the background-subtracted signal reached a minimum signal-to-noise ratio of 120, at which point a new bin was started. Background was subtracted as a single average level per profile.
We computed the PF in each adaptive energy bin using the debiased root-mean-square (RMS) estimator;
\[
PF_{\rm RMS} = \frac{\sqrt{\frac{1}{N}\sum_{i=1}^{N}(r_i - \langle r \rangle)^2 - \sigma^2_{\rm noise}}}{\langle r \rangle},
\]
where $r_i$ are the background-subtracted intensities in each of the N=32 phase bins, $\langle r \rangle$ is the mean intensity, and $N$ is the number of phase bins.
$\sigma_{\rm noise}^2$ is the expected Poisson-noise contribution to the observed bin-to-bin variance, propagated from the source and scaled background count rates.
Uncertainties on both estimators were obtained from 200 Poisson resamplings of the folded source and background profiles. 

Fig.~\ref{fig:pf_heatmap} shows the resulting fine-binned pulsed-fraction evolution alongside the corresponding energy-phase pulse profile, obtained on the same adaptive energy grid described above. The upper panel shows the normalized pulse intensity as a function of energy and phase, with the folded profiles overlaid for reference in each of six broad bands; the lower panel shows the corresponding $PF_{\rm RMS}$ as a function of energy, on the same energy axis.

The $PF_{\rm RMS}$ increases with energy during the first and second \nustar{} observations.  During the first observation, the $PF_{\rm RMS}$ rises sharply after 10 keV, while in the second observation $PF_{\rm RMS}$ again retains a similarly low value in soft X-rays before rising sharply around 30--40 keV (see Fig.~\ref{fig:pf_heatmap}). In contrast, the third observation $PF_{\rm RMS}$ shows an early rise with energy, before reaching a peak value between 5--6 keV, and then it decreases with energy.
Similar negative energy dependence of the pulse fraction has been reported during the decay phase of the 1997 April outburst observed with \textit{BeppoSAX} \citep{2000A&A...357..501P}, at a flux level of $(2.9 \pm 0.3)\times10^{-9}$~erg~cm$^{-2}$~s$^{-1}$ in the 0.3--100~keV band.
Nevertheless, in all observations the $PF_{\rm RMS}$ value at high energies (i.e., 40--78 keV) has similar values around 10\%, and most of the changes seem to occur at the lower part of the pulsed fraction spectrum.

\begin{figure}[]
\centering
\includegraphics[height=4.2in, width=3.4in, angle=0]{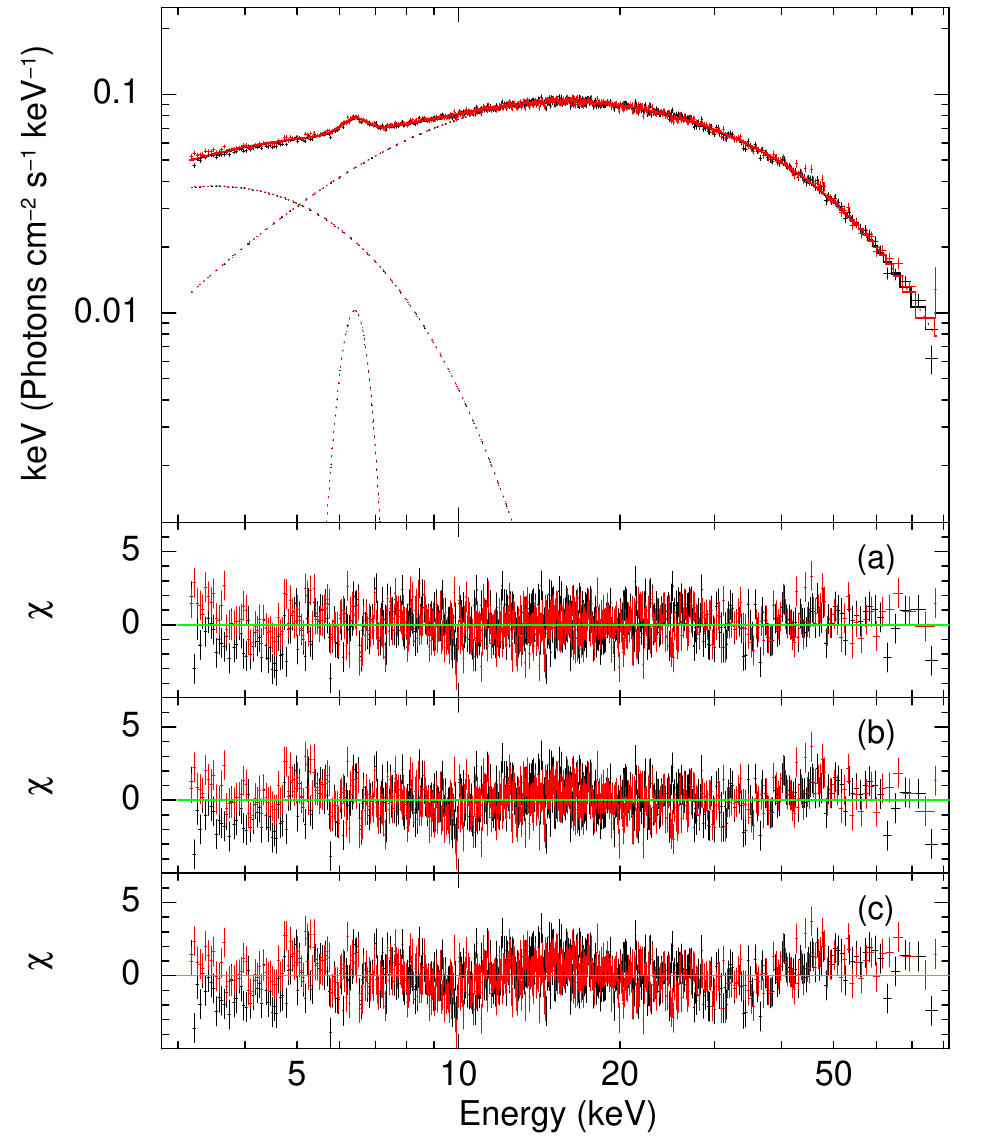}
\caption{Best-fit 3--79~keV energy spectrum from the first \nustar{} observation of GS~1843+00 (top panel). The FPMA and FPMB spectra are shown in black and red, respectively, and are modeled with an absorbed Comptonization model with a blackbody and a 6.4~keV iron emission line. The corresponding residuals from this model are shown in panel (a). Panel (b) corresponds to residuals using an NPEX model with a blackbody and iron line, while panel (c) displays the residuals from a cutoff power-law model with a blackbody and iron line.}
\label{fig:spec}
\end{figure}

\subsection{Broadband spectroscopy with \textit{NuSTAR} and \textit{RXTE}}\label{sec:spec}

To examine the nature of continuum emission from \source, we performed spectroscopy in the 3--79~keV range using \textit{NuSTAR} observations obtained during the 2026 outburst. Spectral analysis was carried out with the {\tt XSPEC} software package (v12.15.1; \citealt{Arnaud1996}). Standard continuum models, including the cutoff power-law \citep{Nagase1989}, the NPEX (Negative and Positive power-law with EXponential cutoff) model \citep{Makishima1999ApJ...525..978M}, and the Comptonization {\tt comptt} model \citep{1994ApJ...434..570T}, were considered in our study. In all cases, an additional blackbody component was required to account for excess residuals in the soft X-ray band. Interstellar absorption was modeled using {\tt tbabs} \citep{Wilms2000} and {\tt vern} cross-sections \citep{Verner96}. Due to the lack of soft X-ray coverage below 3~keV, the absorbing column density is poorly constrained and reaches the lower bound imposed during the fitting. We adopted a lower bound of $9\times10^{21}$~cm$^{-2}$ based on the Galactic column density toward the source \citep{2016A&A...594A.116H}. The resulting lower-bound uncertainty on $N_{\rm H}$ should therefore not be interpreted as evidence for negligible absorption (Table~\ref{table:spec-para}). A Gaussian component was included in the model to describe the Fe~K$\alpha$ emission line near 6.4~keV. We also included a constant multiplicative factor in the model to account for cross-normalization uncertainties between the FPMA and FPMB modules of \textit{NuSTAR}.

Among the tested continuum models, the cutoff power-law with a blackbody component and a Gaussian line provides an inadequate description of the spectrum, mainly due to an absorption-like curvature around 20--30~keV (panel~c of Fig.~\ref{fig:spec}). This absorption curvature is weakly visible when fitting the NPEX model combined with a blackbody and an iron line. In contrast, the {\tt comptt} model combined with a blackbody and the iron line provides a better description of the broadband continuum in the 3--79~keV band, without showing any absorption-like curvature in the hard X-rays. We therefore consider it a suitable model for \source. In {\tt XSPEC} terminology, the model is expressed as {\tt constant*tbabs(comptt+bbodyrad+gaussian)}. The best-fit spectral parameters for \textit{NuSTAR} observations are listed in Table~\ref{table:spec-para}, with uncertainties quoted at the 90\% confidence level. The 3--80~keV unabsorbed flux was estimated using the {\tt cflux} convolution model.  We also studied the 3--30~keV \textit{RXTE}/PCA spectra obtained during the 1997 and 2009 outbursts using the Comptonization-based model. During the 1997 \textit{RXTE} observation, a blackbody component and an Fe~K$\alpha$ emission line near 6.4~keV were required in addition to the continuum model. The resulting best-fit parameters for both \rxte{} observations are listed in Table~\ref{table:spec-para}.

Based on \textit{NuSTAR} observations, we detected a thermal blackbody component with a temperature $kT_{\rm BB}$ of 1.2--1.3~keV. Assuming a distance of 10~kpc, the inferred emitting radius of 2.5--4.2~km is consistent with emission from a region on the neutron star surface. The broadband continuum during the 2026 outburst is well described by a thermal Comptonization model with a seed photon temperature of $T_0 \approx 2.8$--3.4~keV, plasma temperature $\rm Comp_{kT} \approx 10.4$--10.9~keV, and optical depth $\tau \approx$ 3.8--4.2 (Table~\ref{table:spec-para}). In comparison, the \textit{RXTE} spectra from the 1997 and 2009 outbursts indicate a lower plasma temperature ($kT_{\rm e} \approx 6$~keV) and higher optical depth between 6 and 11. No evidence for a CRSF was detected in either \textit{NuSTAR} or \textit{RXTE} spectra.

\begin{figure*}[]
\centering
\includegraphics[height=2.65in, width=4.7in, angle=0]{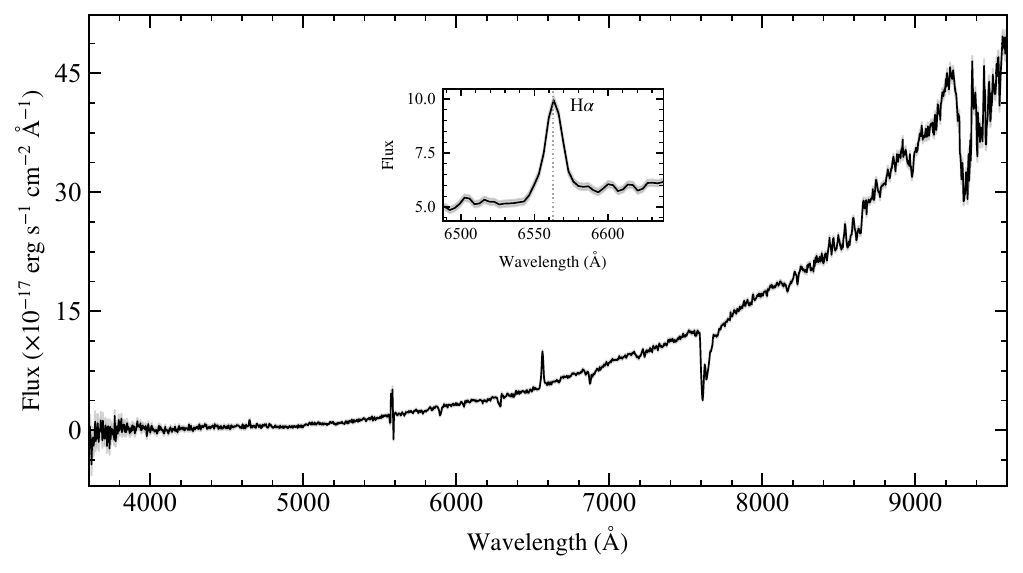}
\includegraphics[height=2.5in, width=6.2in, angle=0]{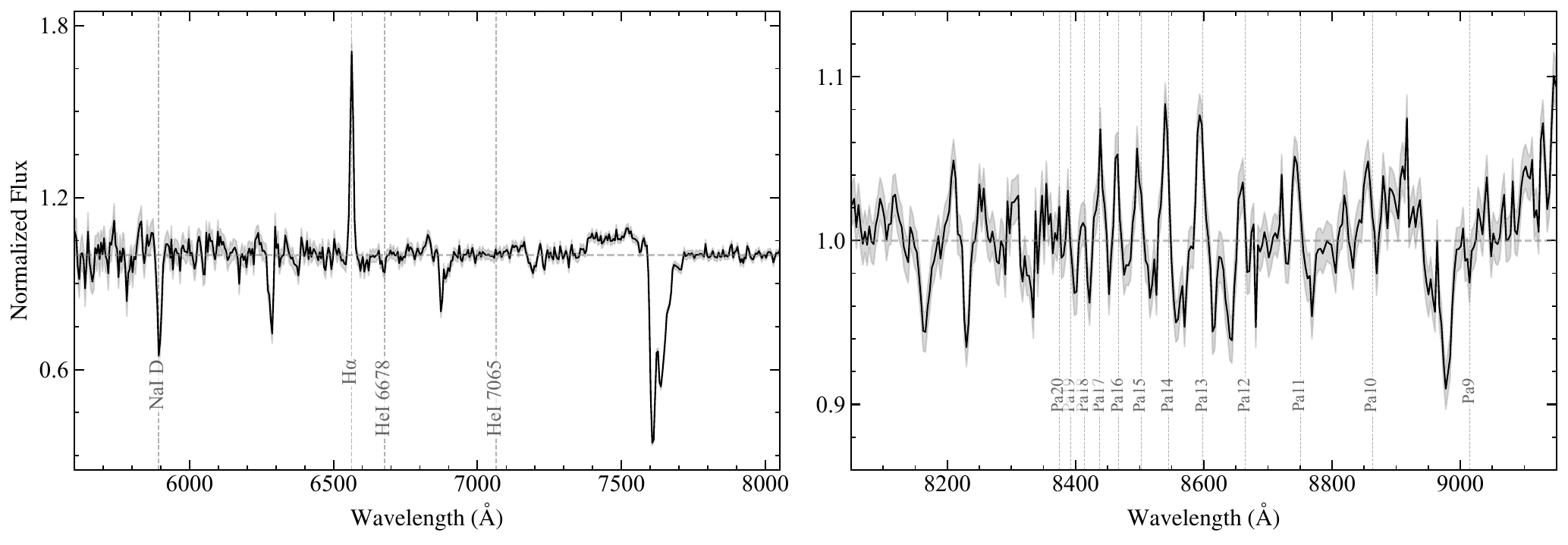}
\caption{Top: optical spectrum of \source obtained using NOT/ALFOSC in 2026 May. H$\rm \alpha$ is detected with weak Paschen series emission lines. Bottom panels show zoomed spectra in optical and near-infrared bands.}
\label{fig:optical_spec}
\end{figure*}

\subsection{Phase-resolved spectroscopy}

Although the first and second \nustar{} observations were taken at similar flux levels around the 2026 outburst peak, their pulse profiles show remarkably different morphologies. To examine the origin of these differences, we performed phase-resolved spectroscopy by extracting spectra in 10 phase bins from both observations. The 3--79~keV spectra from each phase bin were fitted with the absorbed {\tt comptt} continuum with a blackbody component and a Gaussian for the Fe~K$\alpha$ line. All model parameters were allowed to vary during fitting.

Figures~\ref{fig:prs1} and \ref{fig:prs2} show the pulse-phase evolution of the Comptonization parameters (T$_0$, Comp$_{\rm kT}$, $\tau$, and Comp$_{\rm Norm}$),  blackbody parameters (temperature BB$_{\rm kT}$, normalization Norm$_{\rm BB}$, and  blackbody flux Flux$_{\rm BB}$), and the 3--80~keV unabsorbed flux from first and second observations, respectively. The iron line parameters, including energy, width, normalization, and equivalent width, are also shown. The hydrogen column density remains nearly consistent at $<$2$\times$10$^{22}$~cm$^{-2}$ and is therefore not presented in figures. The 3--80~keV unabsorbed flux varies between (5--6.5)$\times$10$^{-10}$~erg~cm$^{-2}$~s$^{-1}$ in the first observation, following the modulation of the pulse profile (first and second panel on left side of Fig.~\ref{fig:prs1}). In the second observation, the flux is slightly lower, (5--5.5)$\times$10$^{-10}$~erg~cm$^{-2}$~s$^{-1}$ (second panel on the left side of Fig.~\ref{fig:prs2}), indicating a modest reduction in luminosity, consistent with the phase-averaged spectroscopy results.

Other parameters, such as the blackbody temperature, its normalization, and the corresponding blackbody flux in the 3--80~keV range, are shown in the second, third, and fourth panels on the right-hand side of Figures~\ref{fig:prs1} and \ref{fig:prs2}. During the first observation, the blackbody temperature exhibits moderate modulation with pulse phase, while the normalization shows a more pronounced variation, indicating changes in the apparent size or visibility of the polar-cap hotspot as the neutron star rotates.  In the second observation, both the blackbody temperature and normalization vary with phase. Although the blackbody temperature remains comparable between the two observations, a slightly higher normalization and corresponding blackbody flux are observed during the second observation, suggesting a modest increase in the emitting area or hotspot size.

The Comptonization parameters also display phase-dependent variations in both observations (third to sixth panels on the left-hand side of Figures~\ref{fig:prs1} and \ref{fig:prs2}). The seed photon temperature ranges from 3.0--3.5~keV during the first observation, with modulation broadly following the pulse profile. A similar, though more complex, multi-peaked behavior of this parameter is observed during the second observation. The electron plasma temperature varies between 11--12~keV with pulse phase in the first observation, whereas a slightly lower plasma temperature of 10--11~keV is found during the second observation. The optical depth $\tau$ varies between 3--4.5 in both observations, with maximum values generally occurring near the minima of the pulse profile.
The Fe~K$\alpha$ emission line is detected in all phase bins of both observations. The centroid energy remains consistent with weakly ionized iron ($E_{\rm Fe} \approx 6.30$--6.45~keV; fifth panel on right side of figures). The equivalent width varies with phase from 0.25--0.40~keV in the first observation and 0.2--0.6~keV in the second observation (sixth panel on right side of figure). The phase-dependent variation of both the line normalization (bottom panel on right side of both figures) and equivalent width (bottom panel on left side of both figures) suggests changes in the observed line emission with pulse phase.

\subsection{Optical and Near-Infrared Observations}

To study the properties of the companion star and its circumstellar environment, we obtained optical and near-infrared spectroscopic and imaging observations of \source{} using the ALFOSC instrument mounted on the NOT. The spectroscopic observations were carried out on 2026 May 23 and 2026 June 19, covering the wavelength range of 3200--9600~\AA. The flux-calibrated spectrum from the 2026 May observation is shown in the top panel of Fig.~\ref{fig:optical_spec}, with the H$\alpha$ emission region shown in the inset. The bottom panels of the figure display the continuum-normalized optical and near-infrared spectra, with the expected atomic transition wavelengths mainly from hydrogen and helium indicated.

The spectrum shows an H$\alpha$ emission line at 6562.8~\AA~ (top panel of Fig.~\ref{fig:optical_spec}). We measure equivalent widths of $-9.4\pm0.3$~\AA\ and $-9.5\pm0.4$~\AA\ from the 2026 May and June observations, respectively. 
The H$\alpha$ line profile was fitted with a single Gaussian model, yielding a full width at half maximum (FWHM) of $13\pm0.7$~\AA, corresponding to a velocity width of $594\pm32$~km~s$^{-1}$. 
However, the resolving power of ALFOSC Grism~\#4 ($R\sim360$) corresponds to a velocity resolution of approximately 830~km~s$^{-1}$. The H$\alpha$ emission is therefore affected by the instrumental resolution, and the observed FWHM may not provide a reliable constraint on the intrinsic velocity structure of the line-emitting region.
No significant emission features from H$\beta$, He~I~6678~\AA, or He~I~7065~\AA\ are seen. A narrow [O~I] sky emission line at 5577~\AA\ is also visible in the spectrum (Fig.~\ref{fig:optical_spec}). In addition to telluric absorption features, an absorption feature consistent with the Na~I~D~doublet at 5892.5~\AA\ is observed, likely originating from the interstellar medium. In the near-infrared region, weak Paschen-series emission lines from Pa~13 to Pa~17 are visible, with equivalent widths ranging from $-0.2$ to $-0.6$~\AA. Because residual fringing becomes increasingly important toward the reddest wavelengths in ALFOSC Grism~\#4\footnote{\url{https://www.not.iac.es/instruments/alfosc/grisms/}}, we regard the measured equivalent widths of Paschen features as approximate.

We also obtained optical imaging observations of the counterpart of GS~1843+00 on 2026 June 20. The optical star is clearly detected at the refined X-ray position from \chandra{} observation (see Fig.~\ref{fig:optical-radio-plot}). The measured magnitudes of the companion are $V=20.8\pm0.7$, $R=18.7\pm0.2$, and $I=16.97\pm0.05$ mag, consistent with values reported by \citet{2001A&A...371.1018I}. The source exhibits a red optical colour, with $V-R\sim2.1$ and $R-I\sim1.7$ mag. For comparison, we inspected the Gaia DR3 photometric measurements of the counterpart (Gaia~DR3~4278536022438800640), which reports $G=18.61\pm0.01$ mag and a red colour of $BP-RP$ =  3.71$\pm$0.12~mag. These colours indicate significant reddening toward GS~1843+00 that may also include a contribution from the Be-star circumstellar disk.

\subsection{Radio observation with \meer}
\label{sec:radio}

We observed \source{} with the MeerKAT telescope during the late declining phase of the 2026 outburst, approximately 20 days after the peak, to search for radio jet emission. The data were reduced following the procedure described in Section~2.4. The final image, produced using Briggs weighting with a robust parameter of $0$, has a synthesized beam of $12\arcsec \times 5\arcsec$. No radio counterpart is detected at the position of \source{}. Two nearby radio sources are visible to the north and south of the target position (see Appendix~\ref{appen:1} and Fig.~\ref{fig:optical-radio-plot}), at angular separations of approximately $15\arcsec$ and $20\arcsec$, respectively.
The local rms noise around the source position is $12 \mu$Jy beam$^{-1}$, corresponding to a $3\sigma$ upper limit of 36~$\mu$Jy~beam$^{-1}$ on the peak flux density. Assuming a flat radio spectrum ($\alpha=0$), this yields upper limits on the 6~GHz radio luminosity of $<6.8\times10^{27}$ and $<2.7\times10^{28}$~erg~s$^{-1}$ for source distances of 5.1 and 10~kpc, respectively.

\section{Discussion}

We present a multiwavelength study of the poorly explored BeXRB system GS~1843+00 during its 2026 X-ray outburst that lasted for about three months. This event represents a giant (Type~II) outburst occurring after a prolonged quiescent interval of 16~years and marks one of the strongest X-ray activity since its discovery. During  \textit{NuSTAR} observations, the pulsar was accreting at an X-ray luminosity exceeding $10^{37}$~erg~s$^{-1}$ (Table~\ref{table:spec-para}). This places GS~1843+00 in the typical luminosity regime of BeXRBs, above the luminosities associated with normal Type-I outbursts but below the most extreme giant outburst events (see, e.g. \citealt{Reig2011, 2018ApJ...863....9W, 2019ApJ...885...18J}). The observed luminosity and extended duration of the activity suggest that the 2026 outburst was a giant outburst driven by a substantial increase in mass transfer from the Be-star circumstellar disk. The spin-up rate of $\dot{P}=(-5.1\pm0.1)\times10^{-8}$~s~s$^{-1}$ is larger than that measured during the 1997 outburst, indicative of angular momentum transfer from the accretion flow to the neutron star. Such enhanced spin-up is expected when angular momentum is transferred through a transient accretion disk.

\subsection{Pulse profile and pulse fraction}
A remarkable feature of the 2026 outburst is the strong evolution of pulse profile morphology and PF with both luminosity and energy. The first and second \textit{NuSTAR} observations were obtained at comparable flux levels, yet they exhibit substantially different pulse shapes and PFs. Similar differences are also observed when comparing observations from the 1997 and 2026 outbursts at comparable luminosities (see, Fig.~\ref{fig:profile_broad}). These results indicate that the observed variability is not solely governed by the mass accretion rate, but is also influenced by changes in the outburst phase and the accretion morphology near the neutron star surface.

The energy-resolved pulse profiles reveal a more complex picture. During the first and second \textit{NuSTAR} observations, the pulse morphology evolves significantly with energy, transitioning from multi-peaked structures at lower energies to more distinct double-peaked profiles at higher energies. In the second observation, we also detect a substantial phase shift between the soft and hard X-ray pulse profiles, with the pulse minimum shifting by approximately half a cycle between the 3--10~keV and 30--79~keV bands. Such energy-dependent phase shifts suggest that the observed emission at different energies may arise from physically distinct regions within the accretion column or from varying contributions of direct and reprocessed emission components. The energy dependence of the pulse morphology may also be related to changes in the angular distribution of the emitted radiation. In strongly magnetized neutron stars, cyclotron resonant scattering can modify the observed pulse profiles through the energy-dependent redistribution of X-ray photons, together with relativistic effects such as light bending and Doppler boosting (see, e.g., \citealt{2014ApJ...781...30N, 2016MNRAS.457.2749J}). However, since the pronounced phase shift is observed only during the second \textit{NuSTAR} observation, while the first observation at a comparable flux level does not exhibit similar behavior, cyclotron resonant scattering alone is unlikely to account for the observed pulse evolution. Instead, changes in the accretion-column structure, viewing geometry, or the relative contributions of different emission components may play a more important role in producing the observed differences in the pulse profiles. During the third observation, the pulse profile exhibits a different energy dependence, with a double-peaked structure at lower energies gradually merging into a broad single peak toward higher energies. This behavior may indicate further changes in the emission geometry during the decay phase of the outburst.

The complex evolution of the PF provides further evidence for changes in the emission from the neutron star (Fig.~\ref{fig:pf_heatmap}). During the first and second observations, the PF broadly increases with energy, implying that the hard X-ray emission is more strongly beamed than the softer emission. However, the third observation exhibits the opposite behaviour above 7~keV, with the PF decreasing with energy. Such a transition from positive to negative energy dependence of the PF within the same outburst is particularly noteworthy and points to substantial changes in the relative contributions of the emission components that drive the observed X-ray flux. 
Although pulse-profile evolution with luminosity has been reported in several X-ray pulsars, including EXO~2030+375 \citep{2017MNRAS.472.3455E,2021JApA...42...33J}, Swift~J0243.6+6124 \citep{2018ApJ...863....9W}, RX J0209.6-7427 \citep{2020MNRAS.494.5350V}, and SMC~X-2 \citep{2023MNRAS.521.3951J}, the pronounced changes observed in GS~1843+00 at similar luminosity levels are intriguing. The negative correlation between pulsed fraction and energy observed during the third \textit{NuSTAR} observation was also reported during the decay phase of the 1997 outburst with \textit{BeppoSAX} \citep{2000A&A...357..501P}. This similarity may indicate that the pulse properties are linked more strongly to the evolutionary stage of the outburst than to the instantaneous luminosity alone.
The observed pulse-profile variation and reversal in the energy dependence of the PF can be interpreted as signatures of a changing accretion geometry. Variations in the accretion-column structure, changes in the illumination pattern of the magnetic poles, or a redistribution between pencil-beam and fan-beam emission components may alter both the pulse morphology and the energy dependence of the PF near the critical luminosity. Our spectroscopic results support this scenario in which the geometry and visibility of the emitting regions evolve throughout the outburst.

\subsection{Spectral properties}

The broadband X-ray spectrum of GS~1843+00 during the 2026 outburst is well described by thermal Comptonization, together with a soft blackbody component and an Fe~K$\alpha$ emission line, as commonly observed in accreting X-ray pulsars \citep[e.g.,][]{Becker2007}.
The blackbody component, with a temperature of $kT_{\rm BB}\approx1.2$--1.3~keV, corresponds to an emitting radius of 2.5--4.2~km for an assumed distance of 10~kpc.  This radius exceeds that expected for a simple polar-cap hotspot, suggesting that the thermal emission originates from an extended heated region, such as an accretion mound or reprocessed emission near the base of the accretion column.
A comparison between the 2026 \textit{NuSTAR} spectra and the \textit{RXTE} observations obtained during the 1997 and 2009 outbursts reveals notable differences in the Comptonizing plasma. While the 2026 outburst is characterized by an electron temperature of $kT_{\rm e}\approx10$--11~keV and an optical depth of $\tau\approx4$, the earlier outbursts exhibited lower plasma temperatures ($kT_{\rm e}\approx 6$ keV) and substantially larger optical depths ($\tau\approx6$--11). These differences may suggest that the accretion column during the 2026 event was hotter and less optically thick, likely reflecting changes in the density and geometry of the accretion flow.

Phase-resolved spectroscopy supports a changing emission geometry. The modulation of the blackbody normalization and Comptonization parameters with pulse phase indicates that different regions of the accretion column and neutron-star surface become visible during rotation.
The stronger variation in blackbody normalization compared to temperature suggests that geometric effects dominate over intrinsic changes in the thermal emission. In addition, the optical depth to reach maximum values near pulse minima may indicate that the observer's line of sight intersects denser regions of the accretion flow when the direct emission from the accretion column is least visible.
The Fe~K$\alpha$ line remains consistent with weakly ionized iron throughout the pulse, while its normalization and equivalent width vary with pulse phase, suggesting an asymmetric reprocessing region.

Although a tentative CRSF near 20~keV was previously suggested from \textit{Ginga} observations \citep{1995PhDT.......215M}, it was not confirmed with \textit{BeppoSAX} \citep{2000A&A...357..501P}. Likewise, we find no significant evidence for a CRSF in either the phase-averaged or phase-resolved \textit{NuSTAR} spectra. Weak residuals around 20--30 keV appear only when empirical cutoff models are used and disappear when the spectrum is fitted with the physical Comptonization model (\texttt{comptt}), indicating that they most likely arise from continuum modeling rather than cyclotron absorption.
Pulsed fraction spectra so far indicated a possibly higher sensitivity to the presence of spectral features, such as CRSFs, than regular spectra \citep[e.g., see recent works by][]{2026A&A...709A..42A, 2026A&A...710A.357M}. In the pulsed fraction spectra, there is some wiggling structure around 20--30 keV, which could be interpreted as an indication of the presence of a CRSF, but further investigation would be required to establish such a connection. 

\begin{figure*}[ht!]
\centering
\includegraphics[height=3.5in, width=6.3in]{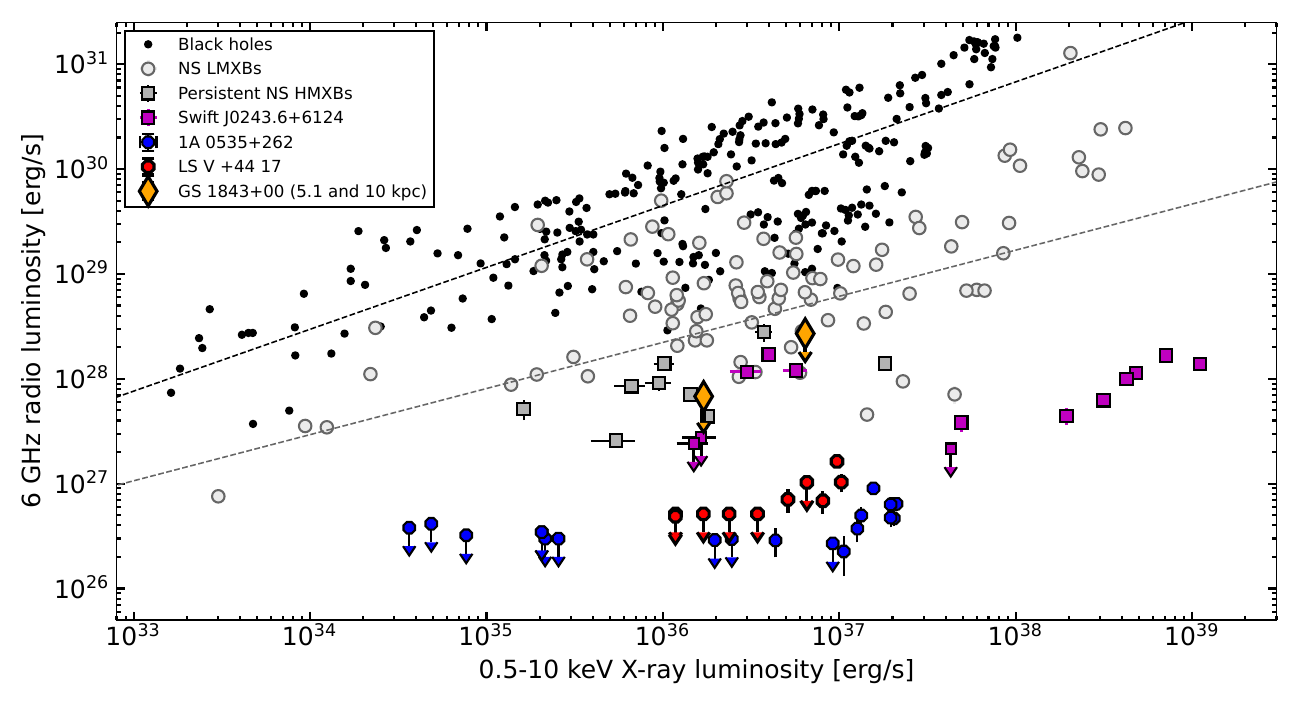}
\caption{X-ray--radio luminosity diagram (adapted from \citealt{2024MNRAS.527.4260V}). The $3\sigma$ radio luminosity upper limits of GS~1843+00 (orange diamonds) are shown at distances of 5.1 and 10 kpc. Black filled circles represent hard-state black hole low-mass X-ray binaries (LMXBs), grey circles indicate hard-state neutron-star LMXBs, and grey squares show persistent high-mass X-ray binaries. Radio measurements of BeXRBs, including Swift~J0243.6+6124, 1A~0535+262, and LS~V~+44~17, are also shown for comparison. The upper limits of GS~1843+00 are consistent with the radio luminosities observed from radio-detected BeXRBs at comparable X-ray luminosities.}
\label{fig:lxlr_plot}
\end{figure*} 

\subsection{Evolution of the Be-circumstellar disk}

The optical spectra obtained with the NOT provide a rare opportunity to investigate the state of the Be-star companion more than two decades after the optical counterpart was identified by \citet{2001A&A...371.1018I}. The detection of H$\alpha$ and weak Paschen-series emission lines confirms the continued presence of a circumstellar decretion disk around the Be star. A comparison with the optical spectroscopy presented by \citet{2001A&A...371.1018I} reveals notable changes in the disk emission properties over this interval. The H$\alpha$ equivalent width measured in our 2026 May and June observations, $EW_{\rm H\alpha}=-9.5\pm0.4$~\AA, is approximately a factor of two smaller than the value of $-18$~\AA\ reported in 2000. In addition, the Paschen-series emission lines are noticeably weaker in our spectra. Since the strengths of the Balmer and Paschen emission lines are generally correlated with the amount of emitting material in Be-star disks (e.g., \citealt{Negueruela1998, Reig2011}), these observations suggest that the circumstellar disk in 2026 contained less emitting material than during the earlier epoch, which may be consistent with a lower-density and/or less extended disk.
The H$\alpha$ emission line in our spectra has a full width corresponding to a velocity of 600~km~s$^{-1}$, somewhat broader than that reported by \citet{2001A&A...371.1018I}.  Although a direct quantitative comparison is limited by differences in spectral resolution and line-profile modeling, we do not use the observed FWHM to constrain the intrinsic velocity structure or radial extent of the line-emitting region. 

The presence of a giant X-ray outburst despite only moderate H$\alpha$ emission highlights the complex relationship between the Be-circumstellar disk and X-ray activity in BeXRBs. While strong H$\alpha$ emission is often associated with a larger or denser disk, the occurrence of Type II outbursts also depends on the disk geometry, warping, truncation, and its interaction with the neutron star, rather than on the total disk mass alone. Since our optical observations were obtained after the peak of the 2026 outburst, it is also possible that part of the circumstellar material had already been depleted through enhanced mass transfer and subsequent accretion onto the neutron star. Alternatively, the outburst may have been triggered by a transient restructuring or warping of the Be disk that temporarily increased the mass-transfer rate without requiring an exceptionally large disk. Continued optical spectroscopic monitoring before, during, and after future X-ray outbursts will be essential for tracing the long-term evolution of the circumstellar disk and establishing its connection with the accretion behaviour of the neutron star.

The optical photometry obtained during the 2026 outburst provides complementary constraints on the state of the Be-star companion. The measured magnitudes, $V=20.8\pm0.7$, $R=18.7\pm0.2$, and $I=16.97\pm0.05$~mag, are consistent, within the uncertainties, with those reported by \citet{2001A&A...371.1018I}. The counterpart exhibits very red observed optical colours, with $V-R\sim2.1$ and $R-I\sim1.7$~mag, consistent with the strong interstellar extinction expected toward GS~1843+00. The \textit{Gaia} DR3 photometry ($G=18.61\pm0.01$~mag and $BP-RP=3.71\pm0.12$~mag) likewise shows a strongly reddened optical counterpart. Given the relatively weak H$\alpha$ emission, the contribution of the circumstellar disk to the broadband optical colours is likely to be modest \citep{2012A&A...539A.114R}. The comparable optical brightness between 2026 and the earlier epoch is therefore not surprising despite the changes in the emission-line strengths. The spectroscopic and photometric observations suggest that the circumstellar environment evolved between the two epochs, while the optical brightness remained largely unchanged. Although the system experienced moderate X-ray activity between 2003 and 2009, the 2026 giant outburst may have been associated with a transient restructuring or redistribution of the circumstellar material that enhanced the mass-transfer rate toward the neutron star.

\subsection{Radio constraints on the jet activity}

The absence of a radio counterpart at the position of \source{} provides a constraint on the presence of transient outflow activity during the late declining phase of the 2026 outburst. Assuming a flat radio spectrum, our $3\sigma$ upper limit of $36~\mu$Jy beam$^{-1}$ at 1.28 GHz corresponds to a luminosity limit of $L_{\rm R}<6.8\times10^{27}$ erg s$^{-1}$ for a distance of 5.1 kpc (and $<2.7\times10^{28}$ erg s$^{-1}$ for 10 kpc). This places \source{} around the radio luminosities measured from the small population of radio-detected neutron-star BeXRBs during their active outburst phases \citep{2021MNRAS.507.3899V, 2024MNRAS.527.4260V}.
Radio emission from these systems has been interpreted as arising from accretion-powered outflows, likely in the form of compact jets, based on the observed association between radio and X-ray activity in systems such as Swift~J0243.6+6124, 1A~0535+262, and LS~V~+44~17 \citep{2021MNRAS.507.3899V, 2024MNRAS.527.4260V}. In these systems, radio emission is typically detected during phases of enhanced accretion luminosity, with the radio brightness decreasing as the X-ray outburst decays. 

Although the MeerKAT and X-ray observations are not simultaneous, the third \textit{NuSTAR} observation was obtained approximately one week earlier at comparable X-ray intensities (Fig.~\ref{fig:maxi-lc}). The resulting radio upper limits for GS~1843+00 are shown in the X-ray--radio luminosity plane in Fig.~\ref{fig:lxlr_plot} for distances of 5.1 and 10 kpc. The upper limits lie within the region occupied by the small sample of radio-detected BeXRBs and are consistent with the luminosities measured from Swift~J0243.6+6124 and other high-mass X-ray binaries at similar X-ray luminosities. The lack of radio detection approximately 20 days after the outburst peak suggests that any radio-emitting outflow had either faded below the MeerKAT sensitivity or remained intrinsically weak during the declining phase. More sensitive, simultaneous radio and X-ray observations spanning the rise, peak, and decay phases of future outbursts will be required to determine whether \source\ launches radio-emitting outflows and how it compares with other BeXRBs in the X-ray--radio luminosity plane.

\section{Conclusions}

We have presented a comprehensive multiwavelength study of the 2026 giant outburst of GS~1843+00. This event, one of the brightest recorded from the source, provides new insights into the accretion physics and circumstellar environment of this poorly explored system. \textit{NuSTAR} observations show that GS~1843+00 reached an X-ray luminosity exceeding $10^{37}$~erg~s$^{-1}$ during the three-month outburst, suggesting its giant (Type II) nature. The measured period derivative indicates substantial net accretion torque during the outburst, likely mediated by a transient accretion disk.

The pulse profiles and pulsed fractions evolve significantly throughout the outburst, even at comparable luminosities. The energy-dependent pulse morphology, including the phase shift between soft and hard X-rays and the reversal of the pulsed-fraction energy dependence during the decay, indicates substantial changes in the emission geometry, likely driven by variations in the accretion-column structure and the relative contributions of pencil- and fan-beam emission.

The broadband X-ray spectrum is well described by a thermal Comptonization continuum with a soft blackbody component and an iron emission line, without any signature of a CRSF. Compared with previous outbursts, the higher electron temperature and lower optical depth suggest a different accretion-column state during the 2026 event. Phase-resolved spectroscopy indicates that the spectral variability is governed by changes in viewing geometry. The \chandra\ observation obtained in 2013 provides an improved X-ray position for the source.

Optical spectroscopy confirms the presence of a Be-circumstellar disk through H$\alpha$ and Paschen emission. However, the weaker H$\alpha$ equivalent width than measured in 2000 suggests a smaller or less dense disk during the 2026 outburst. The occurrence of a giant outburst despite the moderate disk strength implies that enhanced accretion depends not only on disk size but also on its structure, density distribution, and interaction with the neutron-star orbit.

Our MeerKAT observation resulted in a non-detection during the late decline of the outburst. The derived radio luminosity upper limit suggests that any jet or outflow had either faded by the time of the observation or remained intrinsically weak below the detection limit. Radio observations closer to the outburst peak will be crucial for determining whether GS~1843+00 launches transient jets.

In summary, the 2026 outburst of GS~1843+00 reveals a highly dynamic accretion environment in which the neutron-star accretion geometry evolves significantly throughout the outburst. Continued long-term X-ray, optical, and radio monitoring will be essential to establish the connection between Be-disk evolution, mass-transfer episodes, and the accretion behaviour of this system.

\begin{acknowledgements}
We sincerely thank Ignacio Negueruela for carefully reading the manuscript and for his constructive comments and suggestions. We thank the NuSTAR PI for approving our DDT requests, and the NuSTAR SOC for promptly scheduling and executing the observations. This work was supported by NASA through the Astrophysics Explorers Program and made use of data and software provided by the High Energy Astrophysics Science Archive Research Center (HEASARC). Based on observations made with the Nordic Optical Telescope, owned in collaboration by the University of Turku and Aarhus University, and operated jointly by Aarhus University, the University of Turku and the University of Oslo, representing Denmark, Finland and Norway, the University of Iceland and Stockholm University at the Observatorio del Roque de los Muchachos, La Palma, Spain, of the Instituto de Astrofisica de Canarias. The NOT data were obtained under program ID P73-004 (PI Jaisawal).
The MeerKAT telescope is operated by the South African
Radio Astronomy Observatory, which is a facility of the National Research Foundation, an agency of the Department of Science, Technology and Innovation. This work made use of the CARTA (Cube Analysis and Rendering Tool for Astronomy) software (DOI 10.5281/zenodo.3377984 – https://cartavis.github.io). This research also made use of APLpy, an open-source plotting package for Python \citep{aplpy2012, aplpy2019}
\end{acknowledgements}

\bibliographystyle{aa} 
\bibliography{references.bib}

%
%

\begin{appendix}

\onecolumn

\section{Observation log}\label{}

\begin{table*}
\centering
\caption{Log of multi-wavelength observations of \source.}
\label{tab:obslog}
\begin{tabular}{llllll}
\hline
Facility & Instrument/Band & ObsID & Date & Start Time (MJD) & Exposure (ks) \\
\hline

\multicolumn{6}{c}{X-rays} \\
\hline
\textit{RXTE} & PCA & 20143-06-01-00 & 1997-03-05 & 50512.08 & 9.6 \\
\textit{RXTE} & PCA & 94032-11-01-00 & 2009-03-13 & 54903.71 & 3.5 \\
\textit{Chandra} & ACIS & 14644 & 2013-06-09 & 56452.32 & 5.0 \\
\textit{NuSTAR} & FPMA/B & 91201313002 & 2026-03-10 & 61109.94 & 36.0 \\
\textit{NuSTAR} & FPMA/B & 91201319002 & 2026-03-29 & 61128.74 & 37.9 \\
\textit{NuSTAR} & FPMA/B & 91201319004 & 2026-04-16 & 61146.02 & 21.9 \\

\hline
\multicolumn{6}{c}{Optical} \\
NOT & ALFOSC/Grism~\#4  & -- & 2026-05-23 & 61183.068 & 5.10 \\
NOT & ALFOSC/Grism~\#4  & -- & 2026-06-19 & 61210.997 & 5.40 \\
NOT & ALFOSC V-band  & -- & 2026-06-20 & 61211.062 & 0.36 \\
NOT & ALFOSC R-band  & -- & 2026-06-20 & 61211.067 & 0.20 \\
NOT & ALFOSC I-band  & -- & 2026-06-20 & 61211.069 & 0.10 \\

\hline
\multicolumn{6}{c}{Radio} \\
MeerKAT & L-band (1.28 GHz) & -- & 2026-04-23 & 61153.08 & 3.6 \\

\hline
\end{tabular}\label{table:log}
\end{table*}

\newpage
\onecolumn

\section{Lomb--Scargle periodograms}
\begin{figure*}
\centering
\includegraphics[height=2.in, width=2.35in, angle=0]{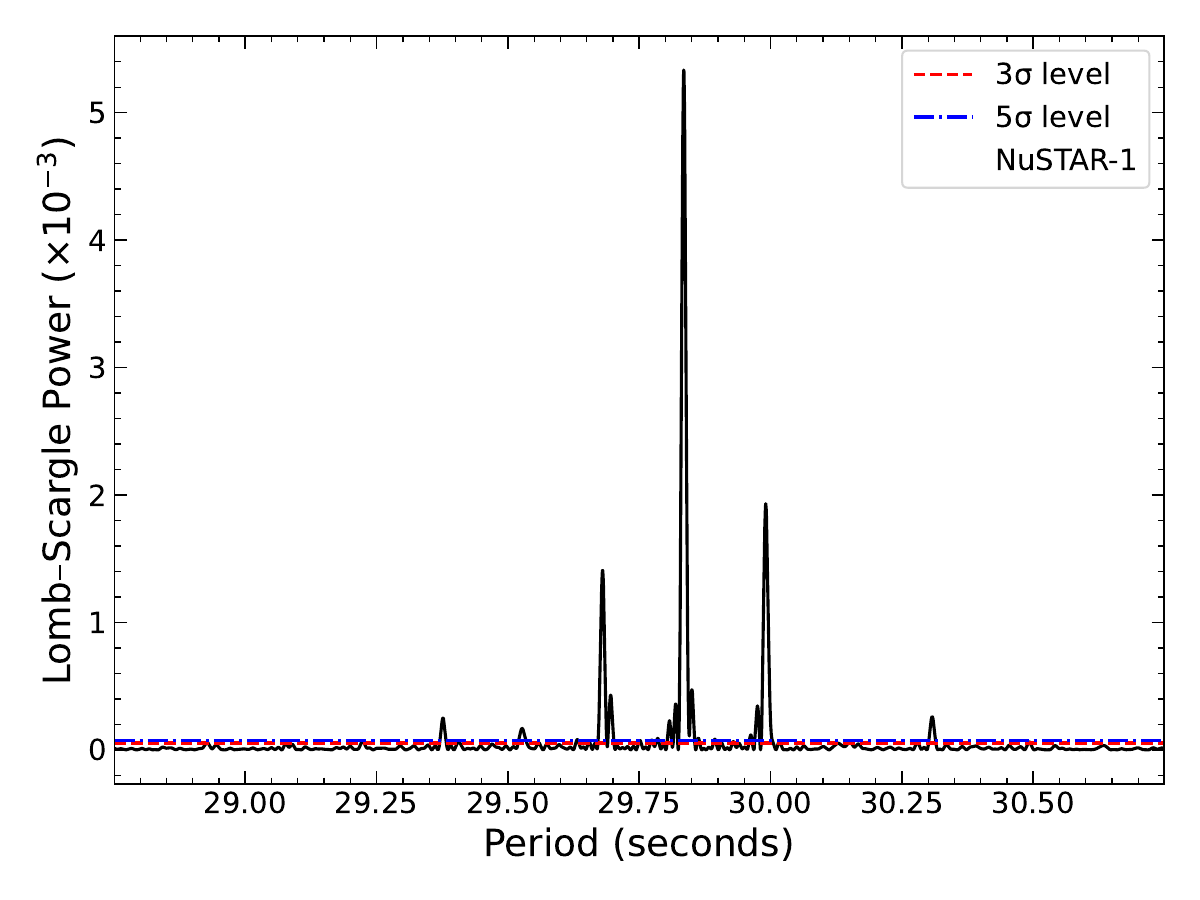} 
\includegraphics[height=2.in, width=2.35in, angle=0]{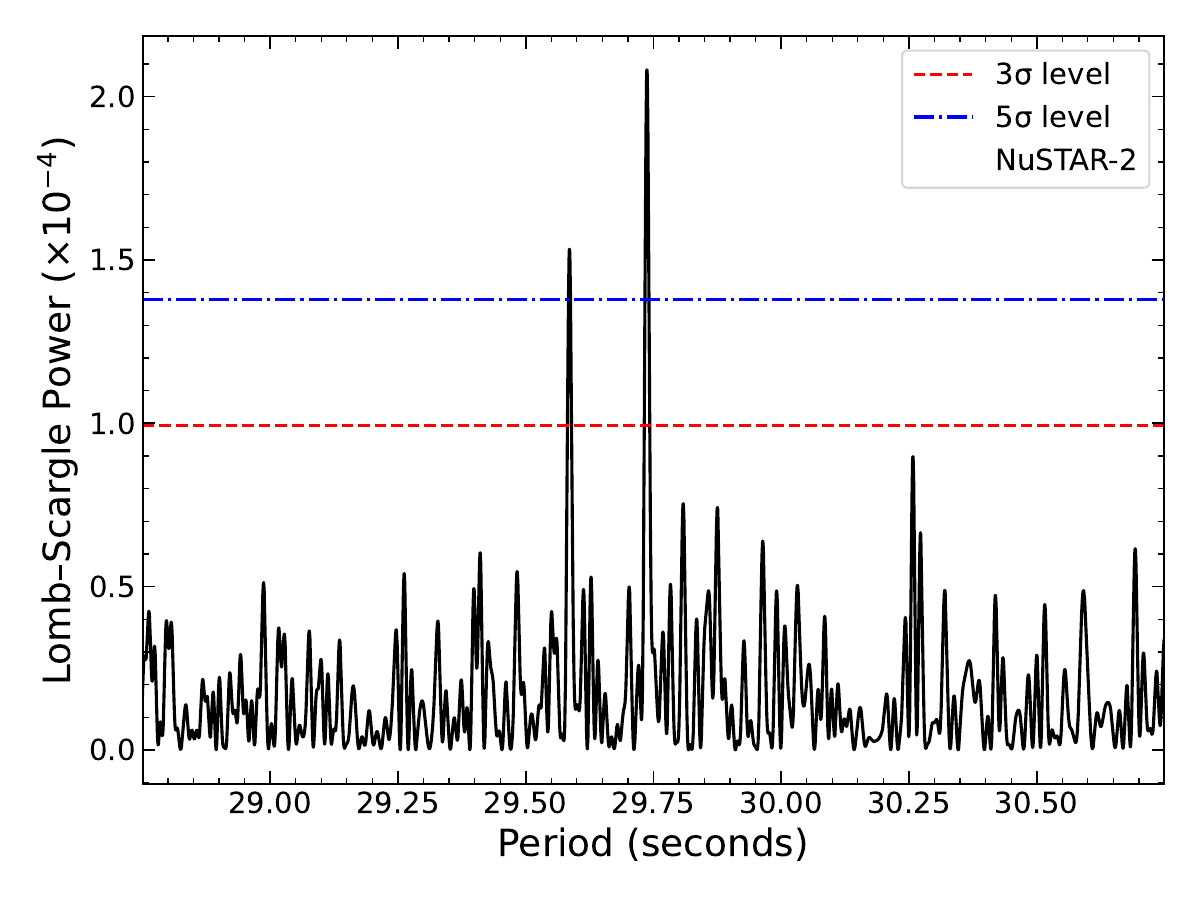} 
\includegraphics[height=2.in, width=2.35in, angle=0]{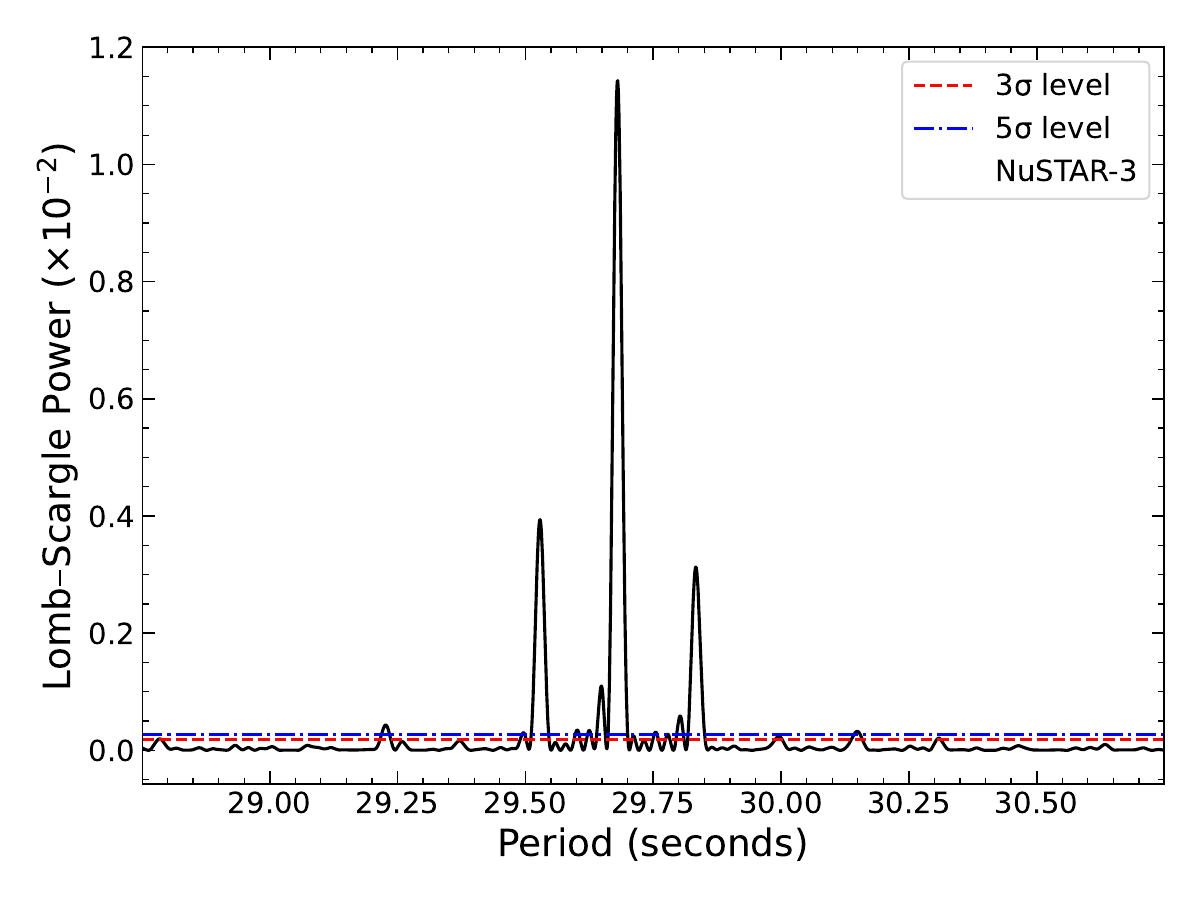}
\caption{Lomb--Scargle periodograms from the first (top), second (middle), and third (bottom) \textit{NuSTAR} observations. X-ray pulsations are detected around 29.7~s. The pulsation signal is weaker in the second observation and increases again during the third observation.  Horizontal lines in each panel indicate the 3$\sigma$ (red) and 5$\sigma$ (blue) levels. }
\label{fig:lspower}
\end{figure*}

\newpage

\section{Phase-resolved spectroscopy} 
\begin{figure*}
\centering
\includegraphics[height=4.8in, width=3.3in, angle=0]{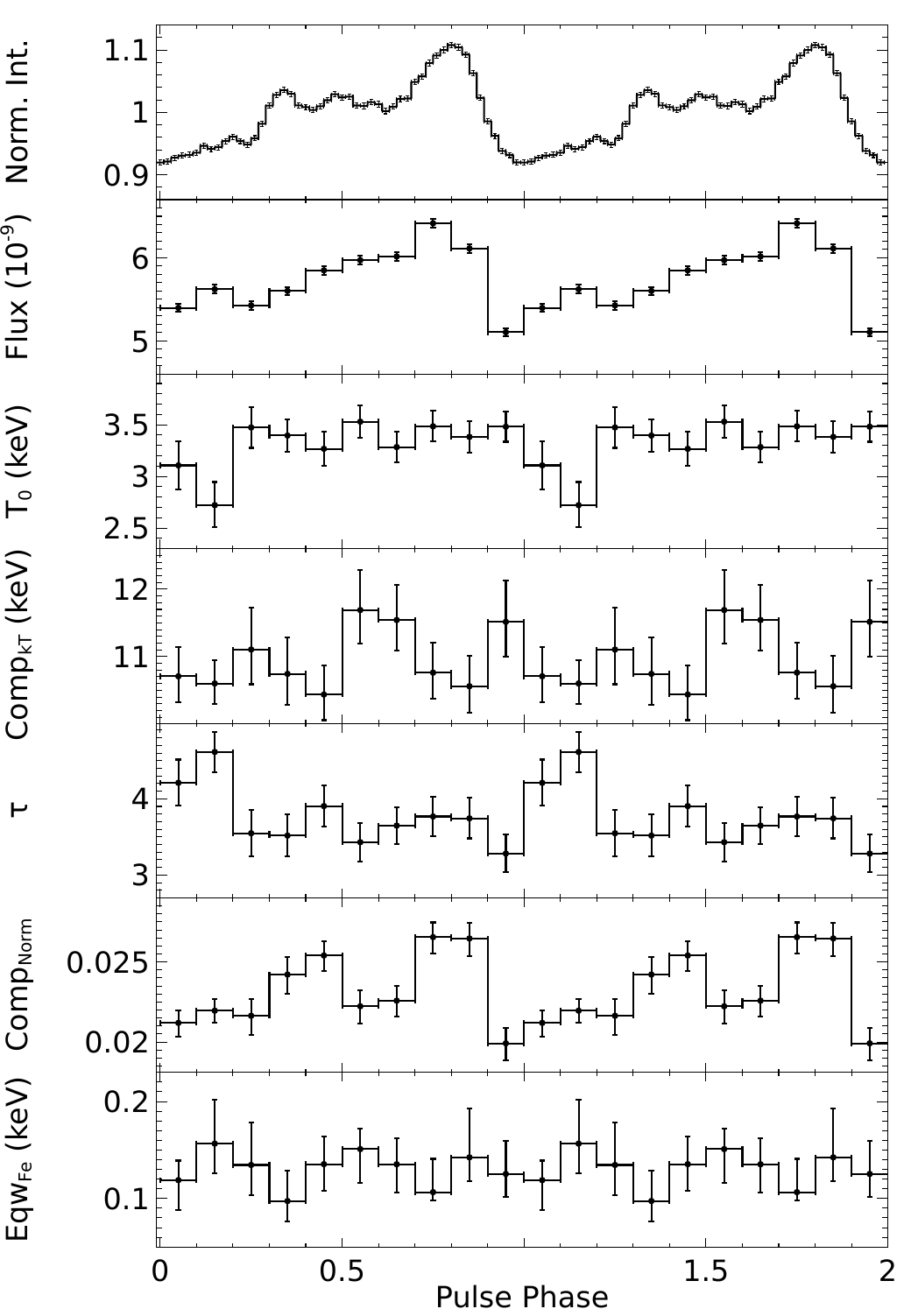}
\includegraphics[height=4.8in, width=3.3in, angle=0]{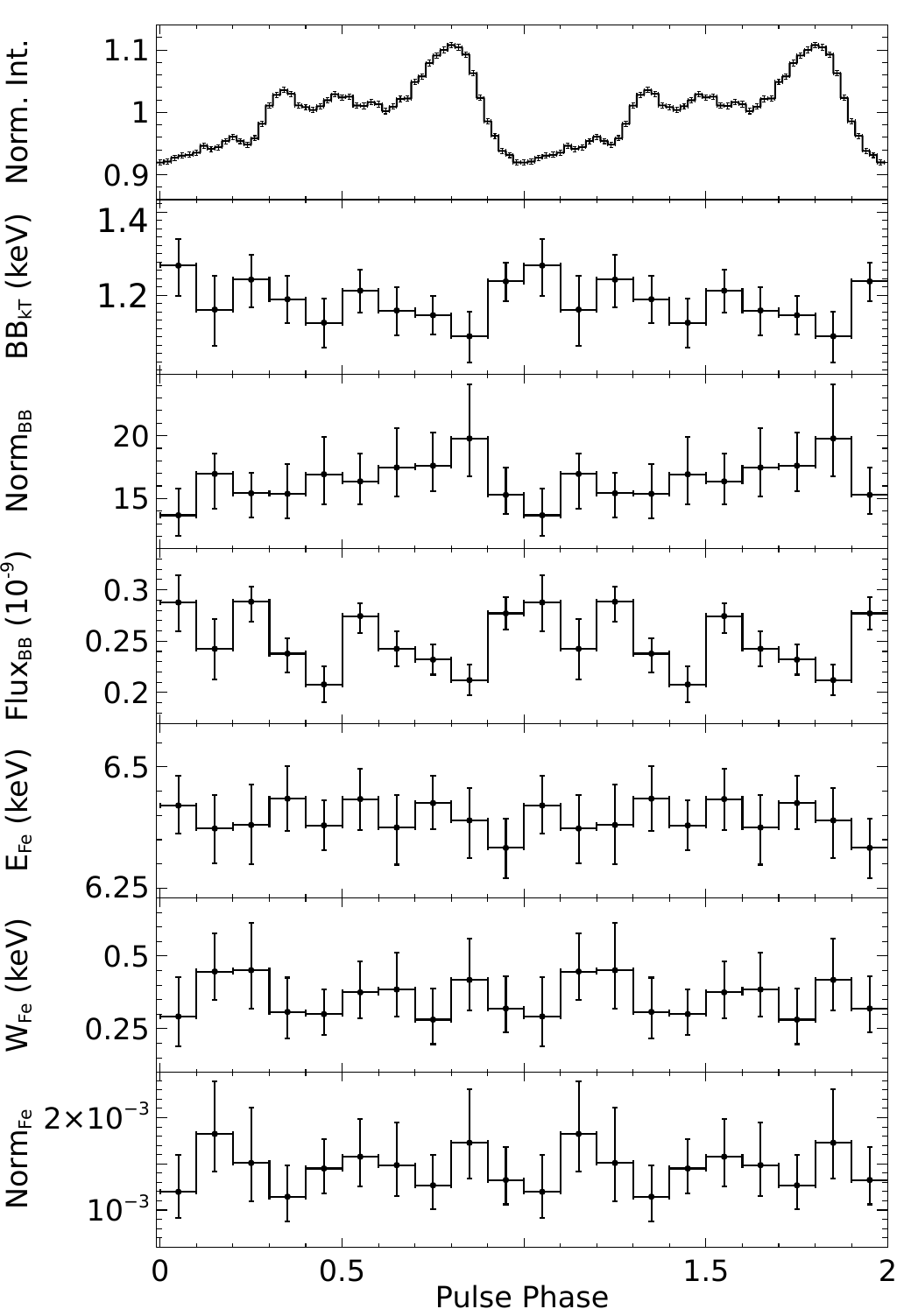}
\caption{Pulse-phase evolution of spectral parameters obtained from phase-resolved spectroscopy of GS~1843+00 using the first \nustar{} observation during the 2026 outburst. The figure shows the variation of Comptonization parameters (seed photon temperature T$_0$, plasma temperature Comp$_{\rm kT}$, optical depth $\tau$, and normalization Comp$_{\rm Norm}$), blackbody parameters (temperature BB$_{\rm kT}$, normalization Norm$_{\rm BB}$, and flux Flux$_{\rm BB}$), iron line parameters (centroid energy E$_{\rm Fe}$, width W$_{\rm Fe}$, normalization Norm$_{\rm Fe}$, and equivalent width Eqw$_{\rm Fe}$), and the 3--80~keV unabsorbed flux (Flux, in units of erg~cm$^{-2}$~s$^{-1}$) as a function of pulse phase. The corresponding pulse profile is shown in the top panel.}

\label{fig:prs1}
\end{figure*}
\begin{figure*}[ht!]
\centering
\includegraphics[height=5.1in, width=3.3in, angle=0]{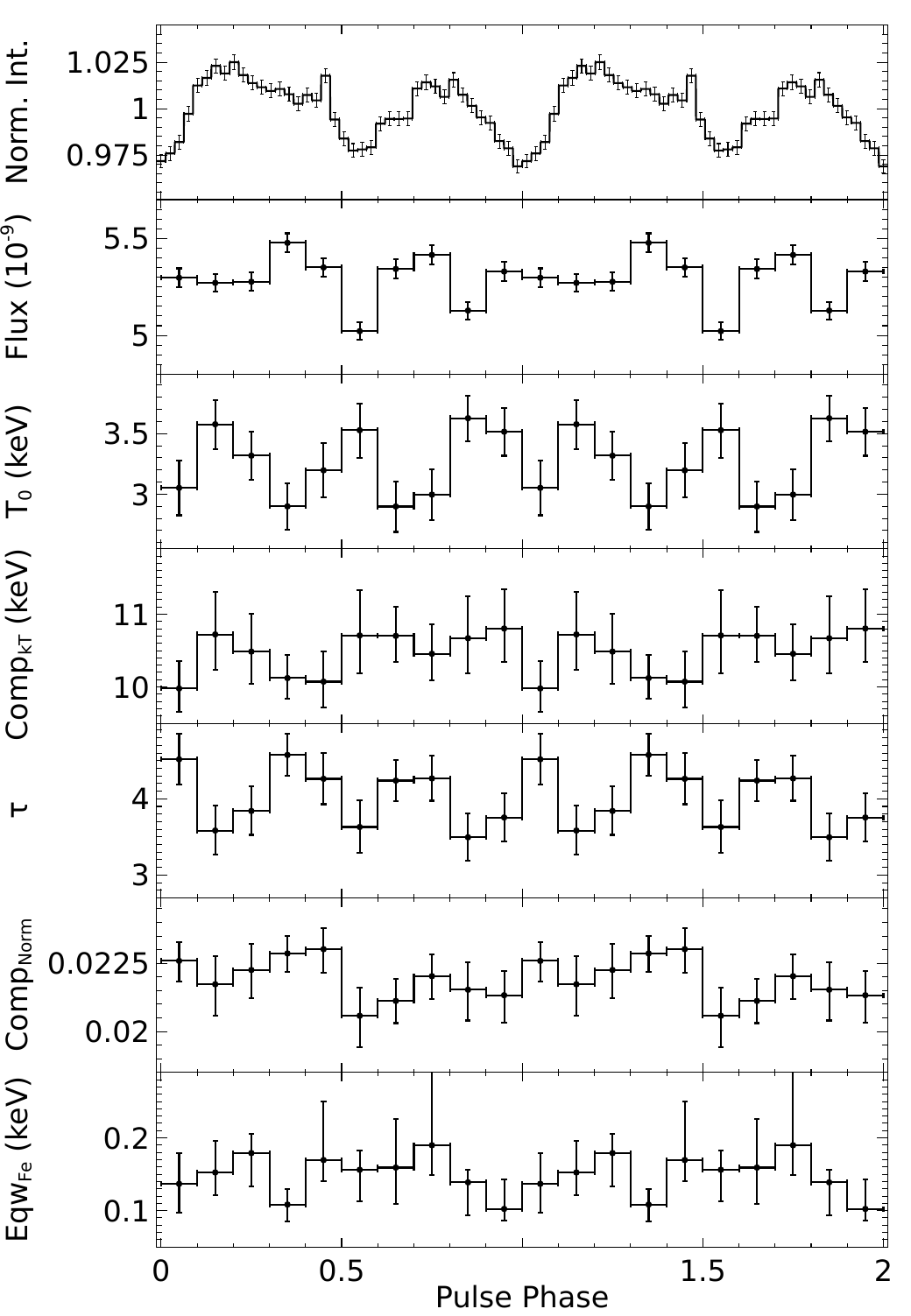}
\includegraphics[height=5.1in, width=3.3in, angle=0]{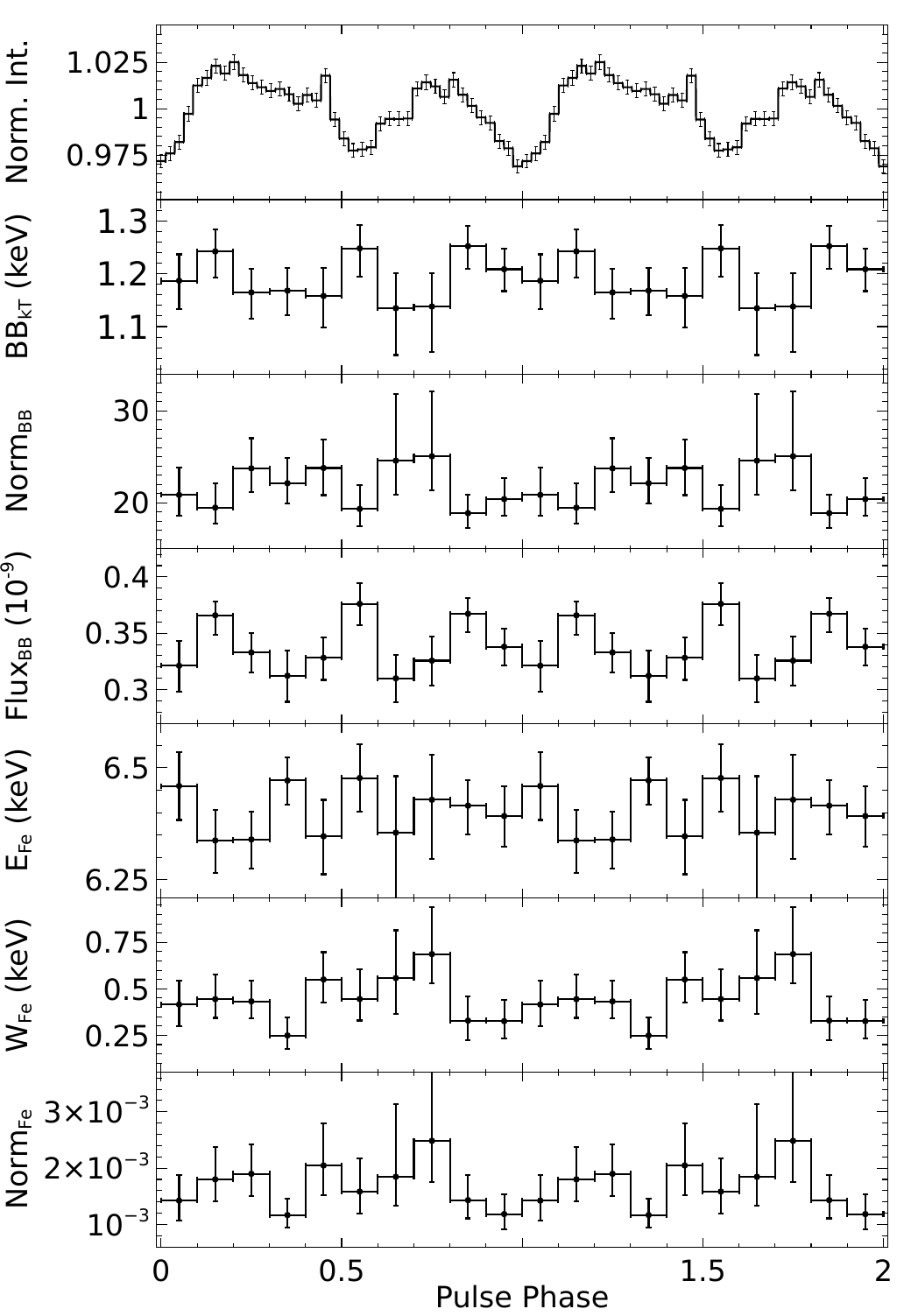}
\caption{Same as Fig.~\ref{fig:prs1}, but for the second \nustar{} observation.}
\label{fig:prs2}
\end{figure*}
\onecolumn

\section{Optical and radio images}\label{appen:1}
\begin{figure*}[h!]
\centering
\includegraphics[height=4.5in, width=5.3in]{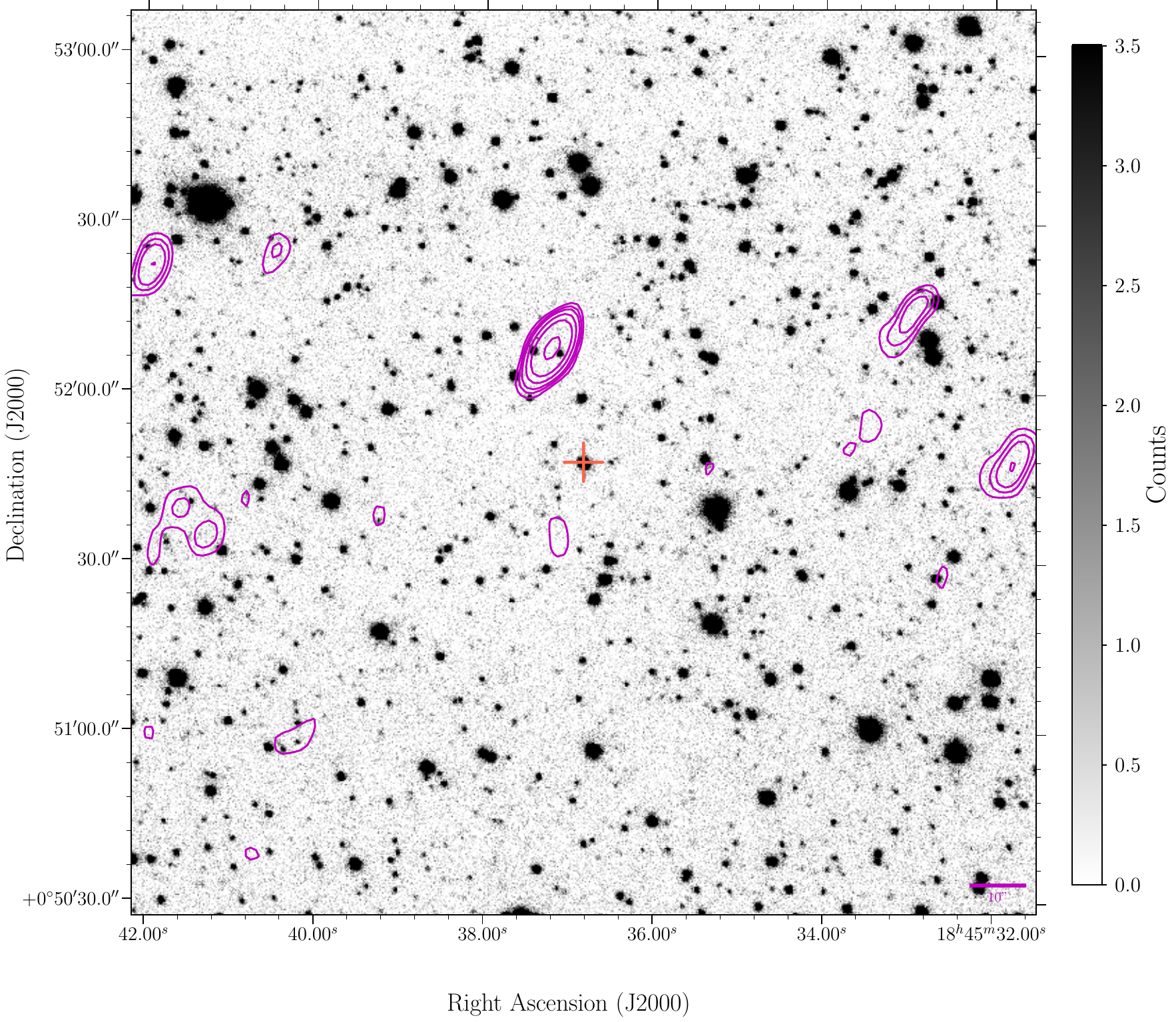}
\caption{ R-band image of the field of GS~1843+00 from NOT/ALFOSC, overlayed with MeerKAT radio contours (magenta) at 1.28 GHz. Contours start at three times the rms (12~$\mu$Jy beam$^{-1}$). The \chandra{} position of GS~1843+00, detected at optical wavelengths but not in the radio band, is marked with a red cross.}
\label{fig:optical-radio-plot}
\end{figure*}

We present the R-band image of the GS~1843+00 field obtained with NOT/ALFOSC in Fig.~\ref{fig:optical-radio-plot}. The refined X-ray position of \source is marked with a red plus sign, corresponding to the position of the optical counterpart. The MeerKAT 1.28~GHz radio contours (magenta) are also overlaid on the optical image, with contours starting at $3\sigma$ above the rms noise level ($\approx12~\mu$Jy beam$^{-1}$).
As discussed in Section~\ref{sec:radio}, two radio point sources are detected in the vicinity of GS~1843+00 with MeerKAT. These sources are located to the north and south of the optical counterpart, respectively. We fitted the radio sources using the CASA {\tt imfit} task. The northern source has a 1.28~GHz flux density of $246\pm9~\mu$Jy, with a position of R.A. (J2000) $=18^{\rm h}45^{\rm m}37.230^{\rm s}\pm0.008^{\rm s}$ and Dec. (J2000) $=00^\circ52'07.6''\pm0.2''$. The southern source has a 1.28~GHz flux density of $53\pm6~\mu$Jy, with a position of R.A. (J2000) $=18^{\rm h}45^{\rm m}37.13^{\rm s}\pm0.02^{\rm s}$ and Dec. (J2000) $=00^\circ51'34.5''\pm0.7''$.

\end{appendix}

\end{document}